\documentclass[twocolumn]{aastex63}

\pdfoutput=1

\usepackage{amssymb}
\usepackage{amsmath}
\usepackage{amsfonts}
\usepackage{graphicx}
\usepackage{color}
\usepackage{hyperref}
\usepackage[caption=false]{subfig}
\usepackage{mathtools}

\def\lsim{~\rlap{$<$}{\lower 1.0ex\hbox{$\sim$}}}
\def\bsim{~\rlap{$>$}{\lower 1.0ex\hbox{$\sim$}}}

\def\hkmsmpc{\ {\rm km\,s^{-1}\,{\it h}Mpc^{-1}}}

\def\hmpc{\ {\rm {\it h}^{-1}Mpc}}

\def\hmsun{\ {\rm {\it h}}^{-1}M_\odot}
\def\msun{\ {\rm M_\odot}}

\def\cmmm{\ {\rm cm^{-3}}}
\def\ccm{\ {\rm cm^2}}
\def\cmm{\ {\rm cm^{-2}}}
\def\cmmm{\ {\rm cm^{-3}}}

\def\cccmss{\ {\rm cm^3\, s^{-1}}}

\def\eV{\ {\rm eV}}

\def\Kel{\ {\rm K}}

\def\kpc{\ {\rm kpc}}
\def\ergss{\ {\rm erg\, s^{-1}}}
\def\ergsscmm{\ {\rm erg\, s^{-1}\, cm^{-2}}}
\def\sfr{\ {\rm M_\odot\, yr^{-1}}}
\def\sfb{\ {\rm erg\, s^{-1} pc^{-2}}}

\def\apjl{Astrophys.\ J.\ Lett.}

\def\aap{Astron.\ Astrophys.}

\def\apjs{Astrophys.\ J. Supp.}

\newcommand{\Halpha}{\mathrm{H}\alpha}

\newcommand{\HI}{\rm{HI}}
\newcommand{\HII}{\rm{HII}}

\newcommand{\Nc}{\big\langle N_c\big\lvert {\it M}\big\rangle}
\newcommand{\Ns}{\big\langle N_s\big\lvert {\it M}\big\rangle}
\newcommand{\mfp}{\lambda_\text{mfp}}

\allowdisplaybreaks

\usepackage[normalem]{ulem}

\newcommand{\chg}[1]{\textsf{\color{red} {#1}}}

\begin{document}

\title{Spatially-resolved modelling of galactic $\Halpha$ emission for galaxy clustering}

\author[0009-0007-7887-783X]{Ivan Rapoport}
\email{ivanr@campus.technion.ac.il}
\affiliation{Physics Department, Technion -- Israel Institute of Technology, Haifa 3200003, Israel}

\author[0000-0003-2062-8172]{Vincent Desjacques}
\affiliation{Physics Department, Technion -- Israel Institute of Technology, Haifa 3200003, Israel}

\author[0000-0002-2539-2472]{Gabriele Parimbelli}
\affiliation{Institute of Space Sciences (ICE, CSIC), Campus UAB, Carrer de Can Magrans, s/n, 08193 Barcelona, Spain}
\affiliation{SISSA, International School for Advanced Studies, Via Bonomea 265, 34136 Trieste, Italy}

\author[0000-0001-9735-4873]{Ehud Behar}
\affiliation{Physics Department, Technion -- Israel Institute of Technology, Haifa 3200003, Israel}

\author[0000-0002-9745-6228]{Martin Crocce}
\affiliation{Institute of Space Sciences (ICE, CSIC), Campus UAB, Carrer de Can Magrans, s/n, 08193 Barcelona, Spain}

\date{\today}

\begin{abstract}
Near-infrared spectroscopic surveys target high-redshift emission-line galaxies (ELGs) to probe cosmological scenarios. Understanding the clustering properties of ELGs is essential to derive optimal constraints. We present a simple radiative transfer model for spatially resolved galactic $\Halpha$ emission, which includes emission from the warm-hot diffuse interstellar medium. The atomic level populations are in steady-state and computed in the coronal approximation. The model is applied to multiple IllustrisTNG simulations in the redshift range $1\leq z \leq 2$ to produce the luminosity function (LF) and the halo occupation distribution (HOD). Collisional processes account for a significant fraction of $\approx 40\%$ of the total $\Halpha$ luminosity ($L_{\Halpha}$). Our LFs are in reasonable agreement with measurements from $\Halpha$ surveys  if a uniform extinction of $0.3<A_{\Halpha}<0.85$ mag is assumed. Our HOD is consistent with that of the \textit{Euclid} Flagship galaxy mock up to differences that can be attributed to baryonic feedback, which is absent from the latter.
When $\Halpha$ luminosities are computed from an empirical relation between $L_{\Halpha}$ and the total star formation rate (SFR) the resulting LFs are in tension with previous observations.
Our approach can be extended to other atomic lines, which should be helpful for the mining of high-redshift galaxy spectra in forthcoming surveys. 
\end{abstract}

\section{Introduction}
\label{sec:intrp}
Emission-line galaxies (ELGs) are characterized by strong emission lines in their spectra, which can be exploited to obtain accurate redshift determinations. As a result, ELGs provide a (biased) sampling of the matter distribution across large comoving volumes. Measurements of their spatial distribution can be exploited to constrain viable cosmological models \citep[e.g.][]{peebles:1980,kaiser:1987,efstathiou/etal:1990,ballinger/peacock/heavens:1996,peacock/etal:2001,tegmark/etal:2004,cole/etal:2005,eisenstein/etal:2005,guzzo/etal:2008,anderson/etal:2012}. In particular, forthcoming measurement of ELG clustering statistics by ground- and space-based galaxy surveys such as \textit{Euclid}, DESI, and SPHEREx \citep{euclidcollaboration2024,desicollaboration2016,spherexcollaboration2014} will
probe the dark energy equation of state, the initial conditions of the Universe,  the nature of dark matter and the mass of neutrinos with unprecedented accuracy.
In addition to their relevance as tracers of the underlying large scale structure, ELGs have the potential, through observables such as luminosities and line profiles, of revealing the history of galaxy evolution, star formation, and gas dynamics in a crucial epoch coinciding with the peak of galaxy formation.

The NISP instrument on board of the \textit{Euclid} satellite \citep{euclidcollaboration2024,EuclidNISP} will obtain spectra in the wavelength range $0.92\,\mu{\rm m}<\lambda<2.02\,\mu{\rm m}$ and allow the detection of a few thousand $\Halpha$ emission line galaxies per square degree in the redshift range $0.9\leq z\leq 1.8$ \citep{Geach_2010,Pozzetti_2016,Merson_2017}. 
An accurate modelling of the $\Halpha$ emission line is essential to determine the window function of the \textit{Euclid} survey, and maximize the cosmological information that can be extracted from the data. 
$\Halpha$ emission is prominent in HII regions surrounding young massive stars, where it is produced by recombinations of free electrons and protons. For this reason, $\Halpha$ has been used extensively as a tracer of the (instantaneous) star formation rate (SFR) and variability in star formation activity \citep[e.g.][]{Kennicutt_1983, Lee_2009, Weisz_2012, Shivaei_2015}. However, while the connection between $\Halpha$ luminosity and SFR is firmly established, uncertainties remain in the modelling of $\Halpha$ emitters due to additional sources of emission arising from e.g. the hot and diffuse gas in the interstellar medium (ISM). Emission from the ISM is strongly sensitive to the physical state of the gas and, in particular, the electron density and ionization state. The diffuse ISM is expected to contribute between 10\% and 50\% of $\Halpha$ emission \citep[see][for a review]{Kewley_2019_rev}, although narrow-band surveys of local galaxies indicate that it may contribute as much as 70\% to the total galactic $\Halpha$ flux \citep[e.g.][]{Zurita_2000, Kreckel_2016}.

Theoretical approaches to the modelling of $\Halpha$ emitters achieve different levels of self-consistency and complexity. Self-consistent radiative transfer simulations propagate rays of ionizing and Lyman-continuum photons to calculate the emission and transport of $\Halpha$ photons, allowing for the study of dust scattering and escape fractions \citep{Katz_2019,Wilkins_2020,Tacchella_2022}. While being highly detailed, such methods are often computationally too expensive to be applied to a large number of simulated galaxies. Alternative approaches utilize semi-analytical nebular emission models \citep{Pellegrini_2020,hirschmann_2023,Pei_2024} or semi-analytical model of galaxy formation calibrated with observational data \citep{Merson_2017,Nusser_2020,Ortega_Martinez_2024_a,Ortega_Martinez_2024_b} to predict emission line luminosities and cumulative flux counts. 
The model implemented in the Flagship galaxy mock of the \textit{Euclid} collaboration \citep{EuclidFlagship2} relies on a combination of a halo occupation distribution \citep[HOD, see e.g.][]{benson/etal:2000,peacock/smith:2000,scoccimarro/etal:2001,berlind/weinberg:2002,zheng/etal:2005} with empirical relations in order to assign $\Halpha$ luminosities to central and satellite galaxies painted on top of dark matter halos. 

In this work, we present a simple radiative transfer model of galactic $\Halpha$ emission, which can be straightforwardly applied to hydro-dynamical simulations of galaxy formation and evolution. The uniqueness of our approach lies in its balance between a detailed modelling of the physics controlling $\Halpha$ emission, computational tractability and versatility. Our model can be readily extended to other emission lines of the hydrogen atom and heavier elements. Here, we apply our approach to the IllustrisTNG simulations to study the halo occupation distribution of our mock $\Halpha$ emitters assuming the simplest halo mass-based description. 
The HOD framework provides an empirical description of galaxy biasing \citep[for a review,][and references therein]{biasreview} which is routinely applied to either construct mock galaxy catalogues or analyze data \citep[see in particular][for the HOD modelling of the ELGs surveyed by the DESI experiment]{gao2023desi,DESI_ELG_1,DESI_ELG_2}. Therefore, we shall also compare our HOD to the \textit{Euclid} Flagship galaxy mock, and assess the sensitivity of our predictions to model systematics.

The paper is organized as follows: we briefly introduce the IllustrisTNG simulations along with key physical quantities in section \S\ref{sec:TNG}; we describe our $\Halpha$ emission model in \S\ref{sec:Haline}; we present our results for the $\Halpha$ luminosity function and HOD in \S\ref{sec:results}; we conclude in \S\ref{sec:conclusions}.

\section{The Illustris-TNG simulations}
\label{sec:TNG}

\subsection{Specifications}
\label{sec:cosmology}

The IllustrisTNG project (TNG, for short) is a collection of advanced cosmological, magnetohydrodynamical galaxy formation simulations \citep{TNG_1,TNG_2,TNG_3,TNG_4,TNG_5,TNG_6,TNG_7, TNG50a, TNG50b}. 
The simulations evolve a $\Lambda$CDM cosmology from redshift $z=127$ to $z=0$ with different box sizes and resolutions. Galaxy formation is modelled by tracking the coupled evolution of four types of particles or cells: dark matter, cosmic gas, stellar/wind particles, and black holes. The key parameters of the simulations we consider in this work are summarized in Table \ref{tab:cosmological_settings}. While our fiducial simulation is TNG300-1, we will use TNG100-1 and TNG300-3 to illustrate the effect of resolution, and Illustris-1 to illustrate the impact of the galaxy formation model. In addition, we will use a galaxy from the high-resolution TNG50-1 for the sake of visualizing the $\Halpha$ emission. In TNG, baryonic feedback is different from Illustris due to changes in the implementation of black hole growth and galactic winds. In addition, TNG includes magnetism and dual-mode AGN feedback which are absent in Illustris. TNG alleviates some model deficiencies that were present in Illustris with respect to benchmark observations, resulting for instance in more realistic galaxy sizes and morphologies \citep{TNG_7}. The subgrid model parameters of TNG are calibrated on TNG100-1 and were subsequently applied to other TNG simulations, which occasionally results in inconsistencies with 
other volume/resolution runs such as TNG300-1, not reproducing the galactic stellar mass function at the high mass end \citep{TNG_5}. This might impact galaxy counts since a different stellar mass function implies a different star formation history.
\begin{table}[h!]
\hspace{-1.2cm}
\begin{tabular}{|c|c|c|c|}
\hline
\textbf{Parameter} & $m_{\text{dm}}$ & $m_{\text{gas}}$ & $\text{L}$ \\ 
&[$10^{7}\hmsun$]&[$10^{7}\hmsun$]&[$\hmpc$] \\
\hline
$\text{TNG300-1}$ & 3.9 & 0.7 & 205 \\ \hline
$\text{TNG300-3}$ & 254.9 & 47.6 & 205 \\ \hline
$\text{TNG100-1}$ & 0.5 & 0.09 & 75 \\ \hline
$\text{TNG50-1}$ & 0.03 & 0.005 & 35 \\ \hline
$\text{Illustris-1}$ & 0.4 & 0.09 & 75 \\ \hline
\end{tabular}
\caption{Summary of the volumes and mass resolutions adopted in the IllustrisTNG suite of galaxy formation simulations. From left to right, we list the dark matter and average gas particle masses 
and the comoving box length. 
The matter, dark energy and baryonic energy densities as well as the present-day Hubble rate (in units of $100 \hkmsmpc$) adopted for TNG are $\Omega_m=0.3089,\ \Omega_\Lambda =0.6911,\ \Omega_b=0.0486,\ h=0.6774 $, and $\Omega_m=0.2726,\ \Omega_\Lambda =0.7274,\ \Omega_b=0.0456,\ h=0.704$ for Illustris.}
\label{tab:cosmological_settings}
\end{table}

\subsection{Halos and subhalos}
\label{sec:subhalos}

In each snapshot, particles are grouped into halos and subhalos with a two-stage hierarchical search using the \texttt{FoF} (friends-of-friends) and \texttt{SUBFIND} algorithms \citep{davis/etal:1985,springel/etal:2001}, respectively. The \texttt{FoF} algorithm defines parent halos by assigning any two dark matter particles to the same group if their distance is below a fixed linking length of 0.2 times the mean inter-particle spacing. Other types of particles are included in the halos of their nearest dark matter particle. The \texttt{SUBFIND} algorithm classifies further the particles into subhalos by identifying virialized substructures within parent FoF halos. We define an ELG candidate as any subhalo hosting at least 1 gas particle and 1 star particle although, in practice, we will only consider ELGs with a $\Halpha$ line luminosity $\geq 10^{39}\ergss$. As a consequence, subhalo gas particles may extend much farther than the disk up to the boundaries of the dark matter subhalo. Hence, we shall model emission from both the interstellar medium and circumgalactic medium of ELGs.

To compute the halo occupation distribution (HOD) of $\Halpha$ emission line galaxies, we will use the IllustrisTNG flags to determine whether an ELG is a central or a satellite galaxy. In these simulations, the subhalo which has the largest total number of member particles is flagged as the central/primary subhalo of the host parent halo (this does not always result in the most massive subhalo flagged as the central galaxy due to the adaptive spatial discretization of the gas cells). The number of central ELGs residing in a host halo, $N_c$, can be either 0 or 1. By contrast, the number of satellite ELGs in the halo, $N_s$, can be any non-negative integer.

\subsection{Properties of the gas cells}
\label{sec:cells}

We model the $\Halpha$ emission at the level of (adaptive) gas cells using the information provided for each cell about the total mass $m_{\text{gas}}$, mass density $\rho_{\text{gas}}$, the hydrogen fractional mass $X_{\text{H}}$, the electron and neutral hydrogen abundance ratios $x_e,\ x_{\HI}$ and the internal energy per unit mass $u$, the instantaneous star formation rate $\text{SFR}$ and the local flux $f_{\text{AGN}}$ produced by active galactic nuclei. The key physical ingredients of our model are the electron, neutral and ionized hydrogen number densities $n_e$, $n_{\HI}$, $n_{\HII}$, the temperature $T$, the cell volume $V_{\text{gas}}$, as well as SFR and $f_{\text{AGN}}$.  

The volume of the gas cell is computed as
\begin{equation}
    V_{\text{gas}} = \frac{m_{\text{gas}}}{\rho_{\text{gas}}} \;.
\end{equation}
The total hydrogen number density $n_{\text{H}}=n_{\HI}+n_{\text{HII}}$ is computed from
\begin{equation}
    n_{\text{H}}=\frac{X_{\text{H}} \rho_{\text{gas}}}{m_p} \;,
\end{equation}
where $m_p$ is the proton mass. 
The number densities $n_e$, $n_{\HI}$ and $n_{\HII}$ are computed using the abundances as
\begin{align}
    &n_e = x_e n_{\text{H}} \\ & n_{\HI} = x_{\HI}n_{\text{H}} \nonumber\\ & n_{\HII} = (1-x_{\HI})n_{\text{H}} \;. \nonumber
\end{align}
We note that due to the ionization of helium and heavier elements, $x_e$ can be larger than unity. Likewise, we also have $x_e+x_{\HI}\ge1$. 

We write the mean molecular weight as
\begin{equation}
    \mu = \frac{4}{1+3X_{\text{H}}+4X_{\text{H}}x_e} \;,
\end{equation}
where we neglect the mass in metals and electrons. Values of $\mu$ vary from 0.62 for (solar-abundance) fully ionized gas ($x_e \approx 1.2$) to 1.29 for neutral gas ($x_e = 0$). The gas temperature $T$ is then computed assuming the ideal gas equation of state (EOS) 
\begin{equation}
    \label{eq:Tgas}
    T = \frac{(\gamma-1)\mu m_p}{k}\, u \;.
\end{equation}
Here, $k$ is the Boltzmann constant and we adopt an adiabatic index $\gamma=5/3$ appropriate to a monoatomic gas. The thermodynamic temperature inferred above exhibits a floor around $\sim 10^4\Kel$ due to the multi-phase nature of the hydrodynamics model implemented in the Illustris-TNG simulations \citep{Springel_2003}. 
While Eq.~(\ref{eq:Tgas}) correctly represents the temperature of the warm and dilute phase of the interstellar medium (ISM), it is not the correct temperature in regions hosting dense and cold star-forming cores, for which $T\ll 10^4\Kel$. 
We assume a fiducial fraction of $10\%$ for the mass in the warm phase if the cell is star-forming, and $100\%$ otherwise.
This simple prescription accounts for the results of \cite{Springel_2003}. 
This way, $\Halpha$ emission from the warm ISM phase is computed 
uniformly for all the cells from the knowledge of the temperature (through Eq.~\eqref{eq:Tgas}) 
and the rest of the model parameters, except that the emission is reduced to 10\% for star-forming cells. This prescription also takes into account the presence of molecular hydrogen in cold star-forming cores. However, we neglect its presence in the warm-hot phase, as the abundance of molecular hydrogen is negligible at temperature $T\geq 10^4 \Kel$.

\section{Modelling the $\Halpha$ emission line}
\label{sec:Haline}

We consider two sources of $\Halpha$ emission in a galaxy: (i) the warm ISM phase and (ii) star-forming $\HII$ regions. In the warm 
ISM, $\Halpha$ emission is induced by electron collisional excitation (CE) of neutral hydrogen, by photo-excitation (PE) of neutral hydrogen by the combined radiation field of nearby hot stars and local AGN, and by radiative recombination (RR) of protons with electrons. 
We obtain the $\Halpha$ emission rate of the ISM upon solving the set of rate equations for the population $n_i$ of the hydrogen energy levels in steady state ($dn_i/dt=0$). The model includes all $N=25$ bound levels with principal quantum numbers $1 \le n \le 5$, and the ionized level $n_{N+1}$ (a bare proton, i.e., the neutral H continuum).
The rate equation for the $i$th energy level takes the form
\begin{equation} \label{eq:rate_equation_master}
    \frac{dn_i}{dt}=\sum_{j=1}^{N+1}Q_{ij}n_j =0 \;.
\end{equation}
Since ISM densities are low, populations of excited levels are also low, the only depletion of excited levels is through spontaneous emission, and all other transitions from excited levels can be neglected.
The transition rate $Q_{ij}$ under this coronal approximation 
is computed according to the ruleset
\begin{equation}
    \label{eq:Q_ij_master}
      Q_{ij} = \left\{
  \begin{array}{cl}
    -\sum\limits_{k<i} A_{ik}\;, &  \text{if }j=i \\
R_{1i}^{\text{exc}}\;, & \text{if } 1<i\le N \\
    A_{ji}\;, & \text{if } j>i \\ 
R_{\infty i}^{\text{RR}}\;, & \text{if }j = N+1
  \end{array}\right. 
\end{equation}
Here, $A_{ji}$ denotes the Einstein coefficient for $j\to i$ spontaneous emission. $R_{1j}^{\text{exc}}$ is the total rate of CE and PE from the ground level to level i, and 
$R_{\infty i}^{\text{RR}}$ is the rate for RR of a bare proton into level $i$. To have a non-trivial solution to the $N+1$ dependent equations in Eq.~(\ref{eq:rate_equation_master}), an additional normalization is required, for example $n_1=1$. We use the \texttt{CHIANTI} atomic database for all the hydrogen atomic data \citep{Chianti_1,Chianti_2}.

The assumption for a steady state is fulfilled as long as the relaxation timescale of the perturbed ISM toward steady state is shorter than the characteristic evolution timescale of a galaxy. Typical recombination rates in the ISM, assuming an electron density of $n_e=1$\,cm$^{-3}$ are $10^{-14}$\,s$^{-1}$ or higher. Excitation and decay rates are higher. Hence, relaxation times are of the order of millions of years, and the steady state assumption holds.

The total $\Halpha$ photon emission rate per H particle, $R_{\Halpha}$, is then given by the following sum over all possible transitions from levels with principal quantum number $n=3$ to $n=2$
\begin{equation}    \label{eq:emission_rate_master}
    R_{\Halpha} = \sum_{i\in \{n=3\}}\sum_{j\in \{n=2\}}n_iA_{ij}\;.
\end{equation}
In the low-density coronal regime, excitation and recombination can be decoupled. As a result, the total $\Halpha$ luminosity $L_{\Halpha}$ is the sum of the luminosities arising from each physical mechanism,
\begin{equation}
    L_{\Halpha} =L_{\Halpha}^{\text{CE}}+L_{\Halpha}^{\text{PE,stars}}+L_{\Halpha}^{\text{PE,AGN}}+L_{\Halpha}^{\text{RR}}+L_{\Halpha}^{\text{HII}} \;.
\end{equation}
We do not include $\Halpha$ dust extinction and model only intrinsic luminosities. 

\subsection{Collisional excitation in the warm-hot ISM phase}
\label{sec:CE}

The contribution of collisional excitation to the galaxy $\Halpha$ luminosity reads
\begin{equation}
L_{\Halpha}^{\text{CE}}=E_{\Halpha}\sum_{\text{gas cells}}V_{\text{gas}}\,n_{\HI}\,R_{\Halpha}^{\text{CE}}\,\Theta_{\text{SFR}} \;,
\end{equation}
where $E_{\Halpha}\approx 1.89\eV$ is the energy of a resonant $\Halpha$ photon, $R_{\Halpha}^{\text{CE}}$ is the total $\Halpha$ photon emission rate (in s$^{-1}$) per $\HI$ particle due to CE, and the sum is over all the gas particles belonging to the subhalo of interest.
The operator $\Theta_{\text{SFR}}$ encodes the reduction of the collisionally-induced $\Halpha$ emission in star-forming cells,
\begin{equation}
\label{hot_ISM_operator}
  \Theta_{\text{SFR}} = \left\{
  \begin{array}{cl}
    \lambda_\mathrm{h}\;, &  \text{if SFR}>0 \\
    1\;, & \text{if SFR}=0
  \end{array}
  \right.
\end{equation}
where the hot ISM fraction $\lambda_\mathrm{h}=0.1$ is the same for all star-forming cells (see Section \S\ref{sec:cells}).

\begin{figure}[h]
    \centering
    \includegraphics[width=0.49\textwidth]{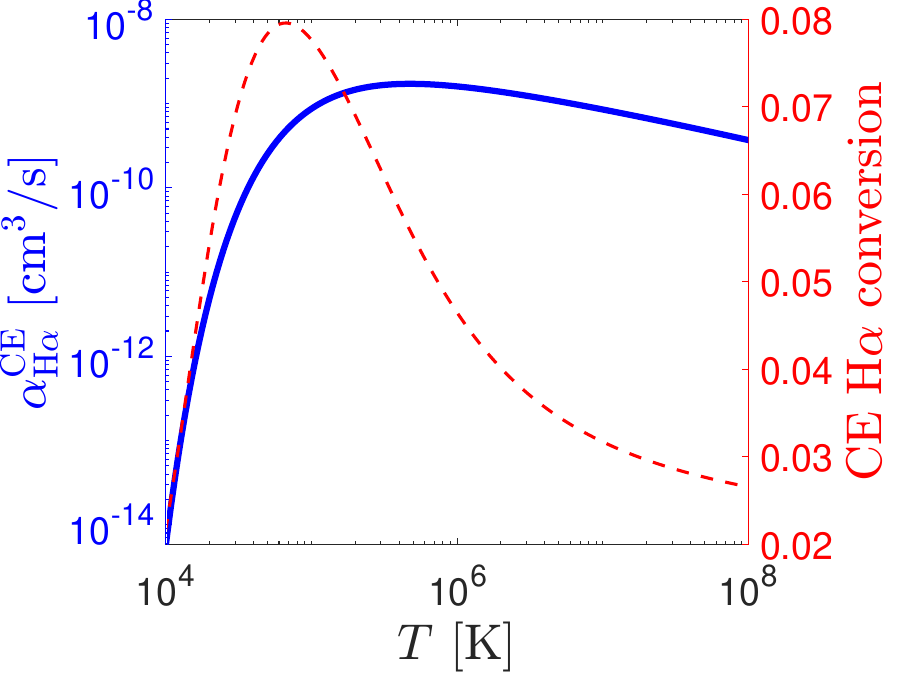}
    \caption{The effective rate coefficient $\alpha_{\Halpha}^{\text{CE}}(T)$ for $\Halpha$ emission from collisional excitations (CE) (solid blue curve), and the conversion probability to produce an $\Halpha$ photon in a CE event (dashed red curve). Both are shown as a function of the gas cell temperature $T$.}
    \label{fig:S_Ha_CE_pdf}
\end{figure}

We adopt the formula of \citet{1992A&A...254..436B} for the rate coefficient $\alpha_{ij}^{\text{CE}}$ for CE from levels $i\to j$ with $E_j>E_i$,
\begin{equation}
\label{CE_Tully_Burgess}
    \alpha_{ij}^{\text{CE}} = 2\pi^{1/2}a_0\hbar m_e^{-1}\left(\frac{I_\infty}{kT}\right)^{1/2}\exp{\left(-\frac{E_{ij}}{kT}\right)}\frac{\Upsilon_{ij}}{g_i} \;.
\end{equation}
Here, $I_\infty \approx 13.6\eV$ is the hydrogen's ionization energy (Rydberg constant), $E_{ij}=|E_j-E_i|$ is the transition energy, $a_0$ is the Bohr radius, $\hbar$ is the reduced Planck constant, $m_e$ is the electron mass, $g_i=2J_i+1$ is the multiplicity of the lower level $i$, and $\Upsilon_{ij}(T)\sim 1$ is the Maxwellian-averaged collision strength. Note also that $2\pi^{1/2}a_0\hbar m_e^{-1}\approx 2.17\times 10^{-8}\cccmss$.
The $i\to j$ excitation rate (per hydrogen atom) is then given by $n_e \alpha_{ij}^{\text{CE}}$. In light of the low ISM density, we neglect collisional de-excitations and set $\alpha_{ij}^{\text{CE}}=0$ if $E_i\geq E_j$. 
In this coronal approximation, the rate equations for $n_i$ reduce to
\begin{equation}
    \label{eq:rate_equation_CE}
    \frac{dn_i}{dt}=n_e\alpha_{1i}^{\text{CE}}n_1+\sum_j \left(A_{ji}n_j-A_{ij}n_i\right) = 0\;,
\end{equation}
where we set $n_1=1$ and compute the emission rate $R_{\Halpha}^{\text{CE}}$ according to Eq.~\eqref{eq:emission_rate_master} from $n_i$ at steady state. It is obvious from Eq.~(\ref{eq:rate_equation_CE}) that the steady state populations satisfy $n_i \propto n_e$, and so does $R_{\Halpha}^{\text{CE}}\propto n_e$. Therefore, we may write
\begin{equation}
\label{R_Halpha_CE}
R_{\Halpha}^{\text{CE}}=n_e\alpha_{\Halpha}^{\text{CE}}(T) \;,
\end{equation}
where $\alpha_{\Halpha}^{\text{CE}}(T)$ is a temperature-dependent effective rate coefficient for the $\Halpha$ emission induced by CE. The solution for $\alpha_{\Halpha}^{\text{CE}}(T)$ is shown in figure \ref{fig:S_Ha_CE_pdf}. The overall $\Halpha$ rate per particle peaks at $T\approx 4\times 10^5 \Kel$, while the conversion probability (defined as the ratio of $R_{\Halpha}^{\text{CE}}$ to the total CE rate) is maximized for temperature $T\approx 0.7\times 10^5 \Kel$ at which approximately 8\% of CE events produce a $\Halpha$ photon. Note that the coronal regime breaks down when the population of the first excited state can grow significantly, which happens when $n_e\alpha_{12}^{\text{CE}}\gtrsim A_{21}$. This, however, implies number densities $n_e\gtrsim 10^8\cmmm$, orders of magnitude larger than those found in the dilute ISM and in illustrisTNG gas cells.

\subsection{Photo-excitation by hot stars}
\label{sec:PE}

The contribution of PE by hot stars to the $\Halpha$ emission in the dilute ISM is computed as
\begin{equation}
L_{\Halpha}^{\text{PE,stars}}=E_{\Halpha}\sum_{\text{gas cells}}\, V_{\text{gas}}\, n_{\HI}R_{\Halpha}^{\text{PE,stars}}\, \Theta_{\text{SFR}} \;,
\end{equation}
where $R_{\Halpha}^{\text{PE,stars}}$ is a rate of $\Halpha$ emission due to PE per $\HI$ atom. 
To compute the photo-excitation rate of the diffuse ISM by stars in each gas cell, we consider first the excitation rate arising from a single source with photon luminosity density $L_E(E)$ (in units of photons erg$^{-1}$ s$^{-1}$). The photon flux decreases with the distance $r$ from this source as
\begin{equation}
 F_E(E) = \frac{L_E(E)}{4\pi r^2} \;.
\end{equation}
The $i\to j$ PE rate $R_{ij}$ induced by a single source is obtained upon integrating over $E$ the previous expression times the PE cross section $\sigma_{ij}(E)$, which exhibits a narrow peak around the transition energy $E_{ij}$. In cgs units, this is
\begin{align}
    \label{R_PE_general}
    R_{ij} &=
    \int F_E(E)\sigma_{ij}(E)dE  \\
    &= \frac{he^2f_{ij}}{4m_ec}\, \frac{L_E(E_{ij})}{r^2} \nonumber \;,
\end{align}
where $e$ is the electron charge and $f_{ij}$ is the oscillator strength of the $i\to j$ absorption. Approximating the spectral energy distribution (SED) of stars by perfect blackbodies, we have 
\begin{equation}
    \label{single_star_LE}
    L_E(E) = \frac{8\pi^2 h R_*^2c^2}{(hc)^4}\frac{E^2}{e^{E/kT_*}-1} \;,
\end{equation}
where $R_*$ is the stellar radius and $T_*$ the effective blackbody temperature. Note that the (bolometric) stellar luminosity recovers the Stefan-Boltzmann law of $L_*=\int EL_E(E)dE=4\pi R_*^2\sigma_\mathrm{SB} T_*^4$ where $\sigma_\mathrm{SB}$ is the Stefan-Boltzmann constant. 

The total PE rate $R_{ij}^{\text{PE,stars}}$ at the locations of interest (e.g. at the center of gas cells) depends on the distances from individual stars, which we do not have from the simulations. To remedy this lack of information, we could generate a Monte-Carlo realization of the stellar spatial distribution consistent with the properties of a given synthetic galaxy at the cell level, and sum $R_{ij}$ over all stars to obtain $R_{ij}^{\text{PE,stars}}$. However, such an approach would be computationally expensive. Therefore, we shall instead approximate $R_{ij}^{\text{PE,stars}}$ by its average over Monte-Carlo realisations of the stellar distribution. This approximation is justified so long as the number of stars per cell is large (which ensures that Poisson fluctuations are small). Furthermore, we shall take into account that radiation is attenuated as it propagates in the hot ISM. 

Summarizing, we express the total PE rate as 
\begin{equation}
    \label{eq:PEstarAVE}
    R_{ij}^{\text{PE,stars}} = \frac{\pi he^2f_{ij}}{m_ec}\, \big\langle J_E(E_{ij})\big\rangle \;,
\end{equation}
where the mean photon specific intensity is defined as
\begin{equation}
\label{eq:JE}
\big\langle J_E(E_{ij})\big\rangle = \left\langle \sum _n \frac{L_{E,n}(E_{ij})}{4\pi r_n^2}\, e^{-r_n/\mfp}\right\rangle \;.
\end{equation}
Here, $L_{E,n}$ is the photon luminosity density of the $n$th star and $r_n$ is its distance from the point of interest. The attenuation length $\mfp$ generally depends on the energy $E_{ij}$ of the photons, the environment of the gas cell etc.
 
To estimate the ensemble average Eq.~(\ref{eq:PEstarAVE}), consider the simplified case of Poisson distributed sources inside a sphere of radius $R$. Assuming a constant $\mfp$ for simplicity and adapting the results of \citet{zuo:1992,2003MNRAS.342.1205M} for the mean specific intensity, the mean photon specific intensity $\langle J_E(E_{ij})\rangle$ at the center of the sphere is given by
\begin{equation}
    \label{eq:meanspecifint}
    \langle J_E(E_{ij})\rangle =  n_* \,\mfp\left(1- e^{-R/\mfp}\right)\, \big\langle L_{E}(E_{ij})\big\rangle 
    \nonumber \;,
\end{equation}
where $\langle L_E(E_{ij})\rangle$ is the average photon luminosity density of stars at energy $E_{ij}$ and $n_*$ is the mean stellar number density. This result provides a useful, and reasonable, approximation to Eq.~(\ref{eq:PEstarAVE}) if the characteristic distance $r_0\equiv \mfp \left(1- e^{-R/\mfp}\right)$ beyond which stars contribute negligibly is not much larger than the characteristic cell size $V_{\text{gas}}^{1/3}\sim \mathcal{O}(1)\kpc$ of the simulations. In this case, stars from the cell under consideration will provide the dominant contribution to $\langle J_E(E_{ij})\rangle$. Consequently,  the ensemble average can be computed locally (i.e. in each gas cell) with the cell-dependent stellar density $n_*$.

To ascertain whether this is the case, consider photons with energy $E_{\text{Ly}\alpha}$ corresponding to the Ly$\alpha$ resonance with cross-section $\sigma_{1s\to 2p}\simeq 10^{-17}\ccm$. We expect that $\mfp$ is of order the average distance between dense gas clouds with column densities $\gtrsim 10^{17}\cmm$. In the reference run TNG300-1, the typical fraction of such high-column density systems in galaxies is a few percents. Ignoring the anisotropy of the spatial distribution of these strong absorbers (they are mainly located in and around the disk), this suggests that $\mfp$ is a fraction of the typical size $R\sim \mathcal{O}(10)$\,kpc of a galaxy, i.e., $\mfp\sim \mathcal{O}(1)\kpc \ll R$. Therefore, we estimate $r_0\simeq\mfp\simeq V_\text{gas}^{1/3}$, which motivates the local approximation 
\begin{equation}
\label{eq:PE_stars}
R_{ij}^{\text{PE,stars}} =\frac{\pi he^2f_{ij}}{m_ec}\, n_*\, r_0\, \big\langle L_{E}(E_{ij})\big\rangle \;,
\end{equation}
for the total PE rate in a given gas cell of stellar density $n_*$. For simplicity, we shall hereafter assume a single, fiducial value of $r_0=1\kpc$ for all cells and galaxies regardless of the photon energy $E_{ij}$. Note that stellar radiation is often attenuated by surrounding neutral gas and dust. This effect translates into a multiplicative escape fraction to $R_{ij}^{\text{PE,stars}}$, which is effectively absorbed into $r_0$. The sensitivity of our predictions to the choice of $r_0$ will be discussed in \S\ref{sec:results}.

Our approximation to the local PE rate depends on the average photon luminosity density $\langle L_{E}(E_{ij})\rangle$ of the stars, which involves an integral over the stellar initial mass function (IMF) $\phi(M_*)$. Our calculations assume a Kroupa IMF \citep{kroupa:2002}
\begin{equation}
    \label{eq:Kroupa_IMF}
      \phi(M_*) = \left\{
    \begin{array}{cl}
    M_*^{-0.3} & (M_*<0.08) \\
    0.08M_*^{-1.3} & (0.08\leq M_* < 0.5) \\
    0.04M_*^{-2.3} & (0.05\leq M_* < 1) \\
    0.04M_*^{-2.7} & (M_*\geq 1)
    \end{array} \right.
\end{equation}
which is normalized to unity for $M_* \in \left(0.05,150\right)M_{\odot}$. Furthermore, we will consider only hot stars with a mass $M_*\geq 15\msun$, which provide the bulk of the PE photons. The short lifetimes of these stars (a few Myr)
allows us to estimate their number density from the knowledge of the instantaneous SFR of each gas cell. Using a unique lifetime $t_*=10{\ {\rm Myr}}$, our final expression for the PE rate of each gas cell is 
\begin{align}
\label{eq:R_PE_stars_final}
    R_{ij}^{\text{PE,stars}} &= \frac{8\pi^3 e^2 f_{ij}E_{ij}^2}{m_e h^2 c^3}\, n_* r_0\\ &\qquad \times\mathcal{N}_{M_*} \int_{15M_\odot}^{150M_\odot} \!\!\frac{R_*^2\,\phi(M_*)}{\exp{E_{ij}/kT_*}-1}dM_* \;,
    \nonumber
\end{align}
where
\begin{equation}
    n_* =  \frac{1}{V_\text{gas}} \int_{15M_\odot}^{150M_\odot} \!\! \frac{t_*\,{\rm SFR}}{M_*}\phi(M_*) dM_*
\end{equation}
is the local, cell-dependent stellar density. The normalization constant is $\mathcal{N}_{M_*}=(\int_{ 15M_\odot}^{150M_\odot}\phi(M_*)dM_*)^{-1}\approx 1.6\times 10^3$ for the IMF considered here. 
To carry out the integration over the IMF, we use the simplified stellar dependencies of $R_*$ and $T_*$ on $M_*$ given by \citet{Demircan1991}
\begin{align}
    &T_*=T_\odot M_*^{0.475} \\ & R_* = R_\odot M_*^{0.8}\nonumber \;,
\end{align}
where $T_\odot = 5785 \Kel$ and $R_\odot = 6.96\cdot 10^{10}\ \text{cm}$.

\begin{figure}
    \centering
    \includegraphics[width=0.49\textwidth]{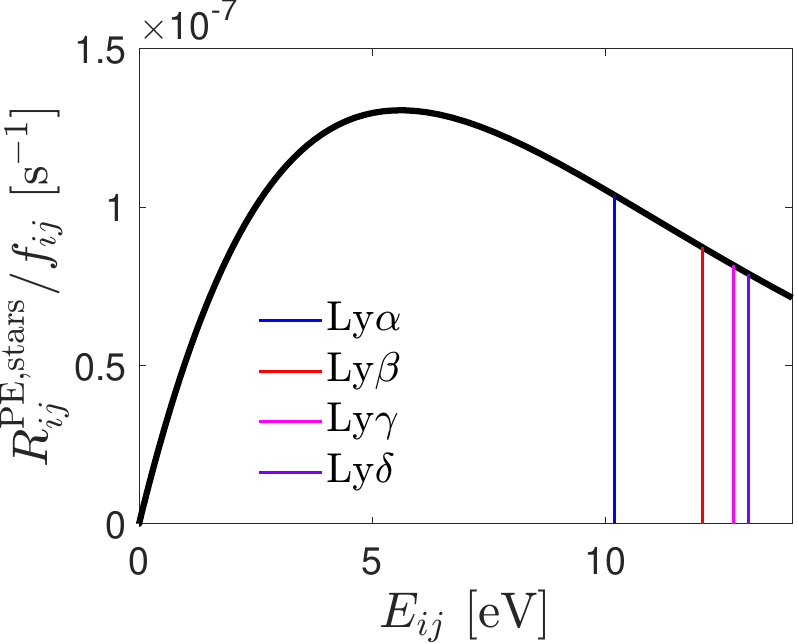}
    \caption{Photo-excitation rate (normalized by $f_{ij}$) as a function of the photon energy $E_{ij}$ for a gas cell with a number density $n_*=10^3{\ {\rm kpc}^{-3}}$ of stars with mass $M_*\geq 15\msun$. The vertical lines mark a few selected transitions $1\to j$ from the Lyman series.}
    \label{fig:PE_R_ij_GF}
\end{figure}

For illustration, the photo-excitation rate~$R_{ij}^{\text{PE,stars}}/f_{ij}$ given by Eq.~(\ref{eq:R_PE_stars_final}) is shown in Fig.~\ref{fig:PE_R_ij_GF} as a function of the photon energy $E_{ij}$ for a gas cell with $n_*=10^3{\ {\rm kpc}^{-3}}$. The vertical lines indicate transitions of the Lyman series as quoted in the figure. We repeat the procedure of \S\ref{sec:CE} and solve
\begin{equation}
    \label{eq:rate_equation_PE}
    \frac{dn_i}{dt}=R_{1i}^{\text{PE,stars}}n_1+\sum_j \left(A_{ji}n_j-A_{ij}n_i\right)
\end{equation}
at steady state to calculate the populations of photo-excited hydrogen atoms. The scaling $R^{\text{PE,stars}}_{1i}\propto n_*r_0$ propagates to the steady state populations, which satisfy also $n_i \propto n_* r_0$. Therefore, the computation of the total $\Halpha$ emission induced by CE events from Eq.~(\ref{eq:emission_rate_master}) can be recast into the form
\begin{equation}
    R_{\Halpha}^{\text{PE,stars}}=\alpha_{\Halpha}^{\text{PE,stars}}\,n_*\, r_0
\end{equation}
where the coefficient
\begin{equation}
    \alpha_{\Halpha}^{\text{PE,stars}}=9.60\times 10^{-13} \ \text{kpc}^2\,\text{s}^{-1}
\end{equation}
does not depend on the properties of the gas cell.
To conclude this Section, we note that including the clustering of the sources would not change the average PE rate in a given gas cell, but it would broaden the variance of fluctuations \citep[see][for a related discussion]{desjacques/etal:2014}. 

\subsection{Photo-excitation by AGNs}
\label{sec:PE_AGN}

Likewise, the contribution of AGN radiation to $\Halpha$ emission in the dilute ISM is
\begin{equation}
    L_{\Halpha}^{\text{PE,AGN}}=E_{\Halpha}\sum_{\text{gas cells}}V_{\text{gas}}\, n_{\HI}\, R_{\Halpha}^{\text{PE,AGN}}\, \Theta_{\text{SFR}} \;,
\end{equation}
where $R_{\Halpha}^{\text{PE,AGN}}$ is the rate of $\Halpha$ emission per $\HI$ atom following a photo-excitation by AGN radiation. To describe the SED of AGNs, we adopt a single power-law with a photon luminosity density index $\Gamma$, such that  $L^{\text{AGN}}_E(E)\propto E^{-\Gamma}$. We select $\Gamma=2$ as the fiducial value. This simple SED is motivated by observations of AGN spectra in both the X-ray and UV energy ranges \citep{Ishibashi_2010, Zheng_1997}. We illustrate the sensitivity of our prediction to the choice of $\Gamma$ in \S\ref{sec:results}. The bolometric luminosity $L_{\rm{AGN}}\equiv \int EL^{\text{AGN}}_E(E)dE$ can be used to write the photon luminosity density as
\begin{equation}
    L_{\rm{AGN}}(E)=\frac{L_{\rm{AGN}}}{\ln(E_+/E_-)}\frac{1}{E^2}=\frac{4\pi r^2f_{\rm{AGN}}}{\ln(E_+/E_-)}\frac{1}{E^2} \;,
\end{equation}
where $E_-=1\eV$ and $E_+=10^5\eV$ are the boundaries of the integration domain and we assume that the AGN radiation is isotropic. The last expression involves the bolometric AGN flux $f_\text{AGN}$, which is evolved in the IllustrisTNG simulations at a cell level. We have checked that our predictions are weakly dependent on the choice of $E_\pm$ owing to the logarithmic dependence. 

In complete analogy with Eq.~\eqref{R_PE_general}, the AGN contribution to the $i\to j$ PE rate is
\begin{equation}
    R_{ij}^{\rm{PE,AGN}}=\frac{\pi h e^2 f_{ij}}{m_e c}\frac{f_{\rm{AGN}}}{(5\ln{10})E_{ij}^2}\exp{\left(-|\Vec{r}_{\text{g}}|/r_0\right)} \;,
\end{equation}
where we have assumed that the resonant AGN radiation is attenuated across the same (uniform) characteristic length $\mfp\simeq r_0 \simeq \mathcal{O}(1)\kpc$ as the stellar radiation, and $\Vec{r}_{\text{g}}$ is the position of the gas cell relative to the galaxy center (defined as the position of the most bound particle in IllustrisTNG). Observe that $R_{ij}^{\rm{PE,AGN}} \propto f_{\text{AGN}}$ regardless of the choice for $L^{\text{AGN}}_E(E)$. 

On repeating the steps outlined in \S\ref{sec:CE} and solving the rate equations
\begin{equation}
    \label{eq:rate_equation_PE}
    \frac{dn_i}{dt}=R_{1i}^{\text{PE,AGN}}n_1+\sum_j \left(A_{ji}n_j-A_{ij}n_i\right)
\end{equation}
at steady state, we arrive at 
\begin{equation}
R_{\Halpha}^{\text{PE,AGN}}=\alpha_{\Halpha}^{\text{PE,AGN}}f_{\text{AGN}}\cdot e^{-|\Vec{r}_{\text{g}}|/r_0} 
\end{equation}
with a (SED-dependent) coefficient
\begin{equation}
    \alpha_{\Halpha}^{\text{PE,AGN}} = 4.45\times 10^{-10}  \ \text{cm}^2\,\text{erg}^{-1} \;.
\end{equation}
The coronal approximation breaks when the rate for resonant Ly$\alpha$ absorption becomes of order the spontaneous, $2\to 1$ decay rate, i.e.  $R_{1{\rm s}\to 2{\rm p}_{1/2}}^{\rm{PE,AGN}}\sim A_{2{\rm p}_{1/2}\to 1{\rm s}}$. Ignoring extinction, this occurs at a critical bolometric flux of $f_{\text{AGN}}\sim 10^{16} \ergsscmm$, which is orders of magnitude larger than the highest flux value $f_{\text{AGN}}\approx 10^7 \ergss \ \text{cm}^{-2}$ among all the gas cells of TNG300-1 at redshift $z=1$.

\subsection{Recombination In The Dilute ISM}

The next physical process we consider is cascade emission from recombination to excited states. The recombination luminosity is computed as
\begin{equation}
    L_{\Halpha}^{\text{RR}}=E_{\Halpha}\sum_{\text{gas cells}}V_{\text{gas}}\, n_{\text{HII}}\, R_{\Halpha}^{\text{RR}}\, \Theta_{\text{SFR}} \;,
\end{equation}
where $R_{\Halpha}^{\text{RR}}$ is the $\Halpha$ emission rate per $\text{HII}$ particle due to recombination events. The corresponding rate equations are
\begin{equation}
    \label{eq:rate_equation_RR}
    \frac{dn_i}{dt}=n_e\alpha_{i}^{\text{RR}}n_{N+1} + \sum_j \left(A_{ji}n_j-A_{ij}n_i\right) \;,
\end{equation}
where, unlike in the previous sections, excited levels are now populated by recombination from the bare state, which we normalize as $n_{N+1} = 1$, and then proceed to relax to the ground state. Since these equations include only recombination transitions and spontaneous emission, the ground level never achieves steady state unlike the excited states. Therefore, we solve Eq.~(\ref{eq:rate_equation_RR}) at steady state for the levels $i\geq 2$ using the case-A recombination rate coefficient of \citet{1932MNRAS..92..820C}
\begin{align}
    \alpha_{i}^{\text{RR}} &= 3.27\times 10^{-12}\left(\frac{10^4\text{K}}{Tn^2}\right)^{3/2}\exp{\left(\frac{\chi_i}{kT}\right)} \\
    &\qquad \times {\rm Ei}\!\left(\frac{\chi_i}{kT}\right)\mathcal S_i\ \cccmss 
    \nonumber \;.
\end{align}
Here, $\chi_i=I_\infty-E_i$ is the energy required to ionize a hydrogen atom in level $i$, $\text{Ei}$ is the exponential integral, $n$ is the principal quantum number of the electron orbital and we have added a multiplicative factor of $\mathcal S_i$ for the following reason. Without $\mathcal S_i$, the above expression provides total rate coefficients as a function of $n$ solely. However, we also need the fine structure in order to solve Eq.~(\ref{eq:rate_equation_RR}). For this purpose, we multiply the original expression of \cite{1932MNRAS..92..820C} by the relative statistical weight $\mathcal{S}_i = g_i / 2n^2$ of the electron orbital with (total angular momentum) quantum number $J_i$ (and multiplicity $g_i = 2J_i+1$). 

\begin{figure}
    \centering
    \includegraphics[width=0.49\textwidth]{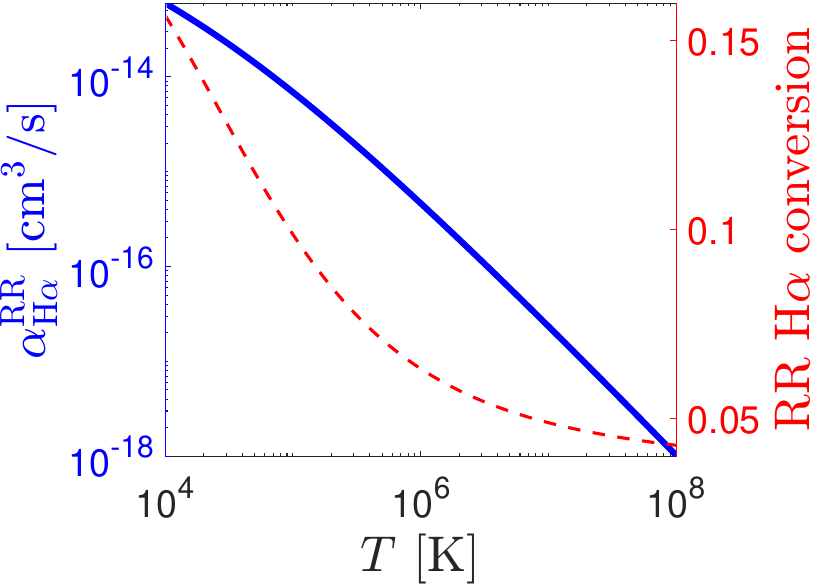}
    \caption{The effective rate coefficient $\alpha_{\Halpha}^{\text{RR}}(T)$ for $\Halpha$ emission from radiative recombination as a function of the gas cell temperature (solid blue curve), and the conversion probability (per recombination event) to produce an $\Halpha$ photon (dashed red curve).}
    \label{S_Ha_rec_pdf}
\end{figure}

Since $\alpha_{i}^{\text{RR}}$ decreases sharply with $n$, an approximation with a finite number of levels above the ground state is sufficient to capture radiative recombinations accurately. Observe also that, like CE-induced emission, the steady state solutions satisfy $n_i\propto n_e$ and $R_{\Halpha}^{\text{RR}}\propto n_e$. Therefore, we can write
\begin{equation}
    R_{\Halpha}^{\text{RR}} = n_e\, \alpha_{\Halpha}^{\text{RR}}(T) \;.
\end{equation}
The function $\alpha_{\Halpha}^{\text{RR}}$ is shown in Fig.~\ref{S_Ha_rec_pdf}. The recombination conversion probability (the probability that a radiative recombination leads to the emission of a $\Halpha$ photon) varies with temperature. It decreases from $\sim 15\%$ at the temperature floor to $\sim 5\%$ at $T=10^8\Kel$.

\begin{figure}
    \centering
    \includegraphics[width=0.49\textwidth]{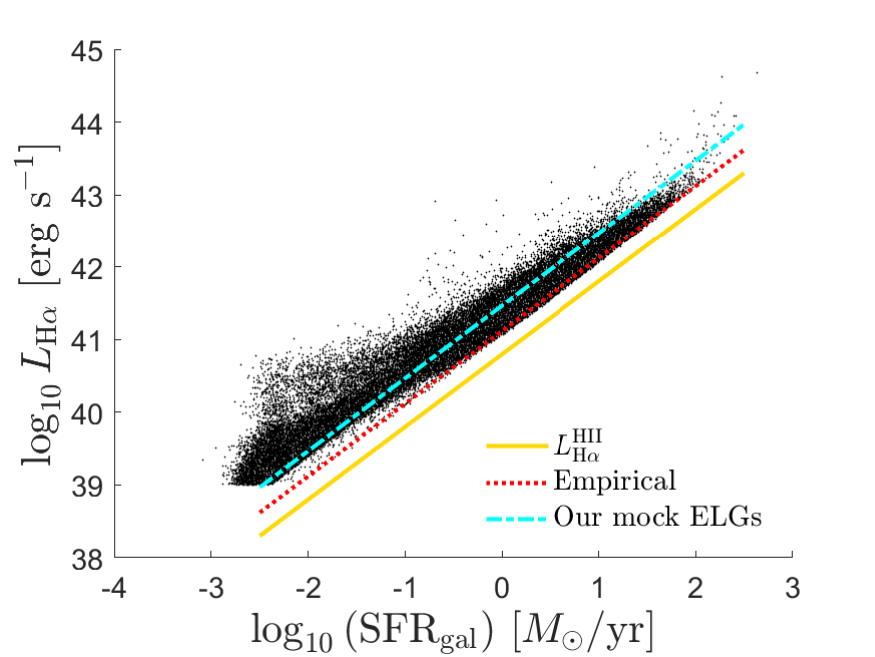}
    \caption{Total galactic SFR ($\rm{SFR_{gal}}$) - $L_{\Halpha}$ relation of all the star-forming ELGs in TNG300-1 ($z=1$). Each black dot of the scatter plot marks an ELG. The yellow line is Eq.~\eqref{eq:L_HII}, the red line represents the empirical relation (cf. main text), and the cyan line displays the best fit linear regression $L_{\Halpha}=10^{41.47}\rm{SFR_{gal}}$ of the scatter plot.}
    \label{fig:HaSFR}
\end{figure}

\subsection{Emission From HII Regions}
\label{sec:HII_regions}
HII regions are typically found in molecular clouds with recent or ongoing star formation, in which massive hot stars ionize the largely cold and neutral hydrogen in their surroundings. $\Halpha$ emission can occur when the ionized gas recombines. Since the physical state of star-forming regions is not properly modeled in the simulations, we take the $\Halpha$ emission from gas cells to be directly proportional to the instantaneous SFR in analogy with the PE-induced $\Halpha$ emission discussed in Section~\S\ref{sec:PE}. Namely, we set 
\begin{align}
\label{eq:L_HII}
L_{\Halpha}^{\text{HII}} &=C_{\HII}\sum_{\text{gas cells}}\, \text{SFR}\hspace{4pt}\, \ergss \\
&= C_{\HII}\, {\rm SFR}_\text{gal} \ergss \nonumber \;,
\end{align}
where ${\rm SFR}_\text{gal}$ is the total star formation rate of an ELG (in unit of $\sfr$). We adopt a constant of proportionality $C_{\HII} = 10^{40.8}$ equal to half the coefficient of the empirical SFR - $L_{\Halpha}$ relation, $L_{\Halpha}=10^{41.1}{\rm SFR}_\text{gal}$. Our choice is motivated by the fact that this relation involves the total $\Halpha$ luminosity, whereas our focus is on using this relation to model only the contribution from HII regions. It is consistent with the fact that $10-70\%$ of the $\Halpha$ emission could arise from the diffuse ISM. This relation was calibrated for $z\lesssim 0.5$ \citep{Kennicutt_1983, Lee_2009,Dominguez_Sanchez_2012, Koyama_2015}, and at $2.08\leq z\leq 2.51$ \citep{Shivaei_2015}. We apply the relation Eq.~(\ref{eq:L_HII}) uniformly to all the galaxies and at all redshifts. 

Fig.~\ref{fig:HaSFR} shows Eq.~(\ref{eq:L_HII}) as the (red) dashed line, whereas the position of each (black) dot indicates the (total) SFR and $L_{\Halpha}$ of a synthetic ELG. $\Halpha$ emission induced by RR, CE and PE enhances the total $L_{\Halpha}$ of a galaxy and generates a significant scatter at fixed SFR. The dashed (cyan) line indicates the best fit linear regression of the sample (after excluding ELGs that are not star-forming, i.e. ${\rm SFR}_\text{gal}=0$, which are about $3\%$ of the full ELG sample). Note also that the fraction of $z=1$ $\Halpha$ emitters with low star formation rate ${\rm SFR}_\text{gal}<1\sfr$ but above the {\it Euclid} flux limit $f_{\Halpha}>2\times 10^{-16} \ergsscmm$ is $\simeq 0.04$\%, and never exceeds $0.1$\% across the redshift range $1<z<2$.

\begin{figure*}[htp]
    \centering
    \includegraphics[width=1.0\textwidth]{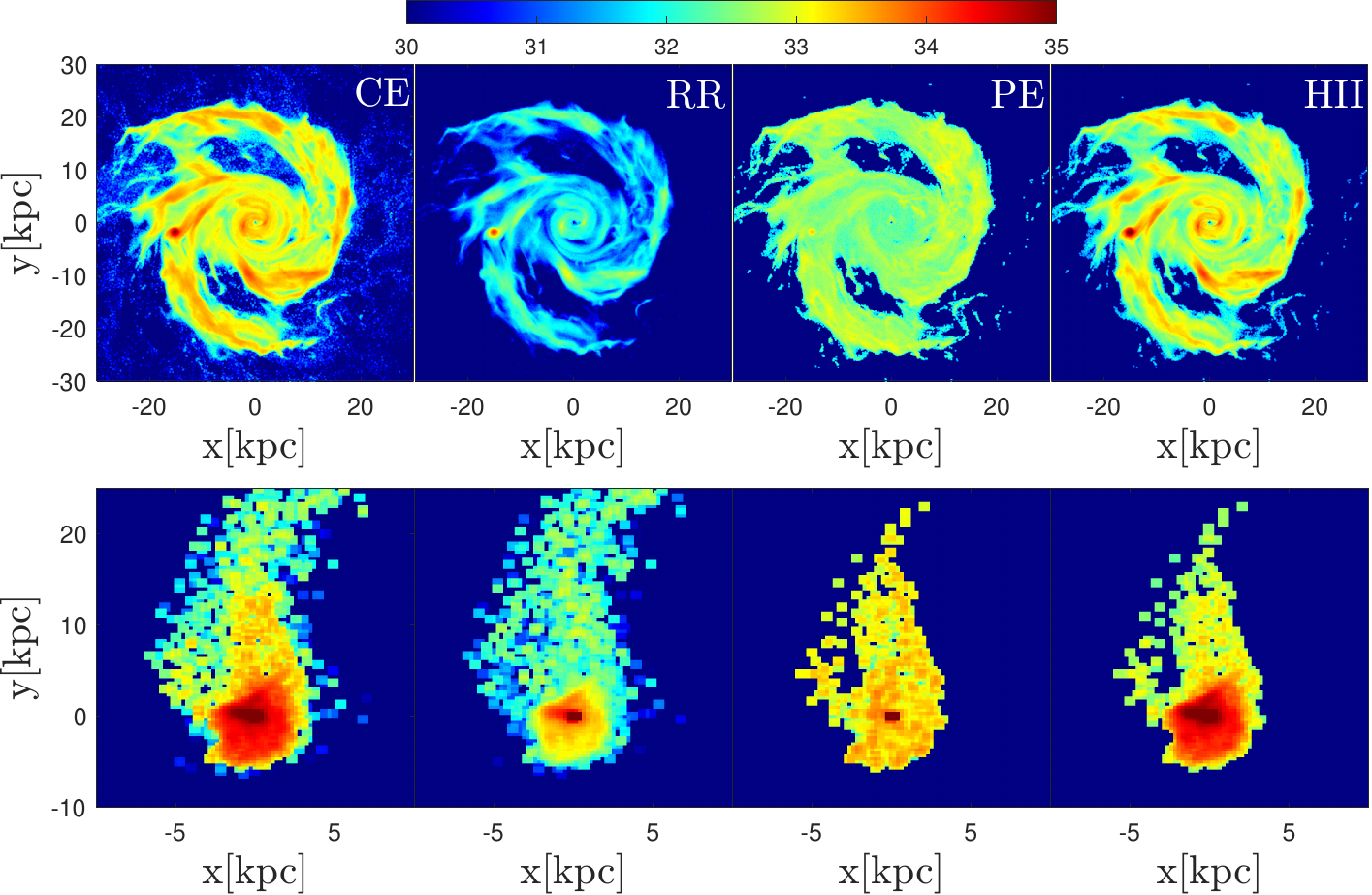}
    \caption{Face-on maps of the $\Halpha$ surface brightness $\Sigma_{\Halpha}$ of two ELGs. The panels from left to right are the separate contributions of collisional excitations, recombinations, photo-excitations and $\HII$ regions. They display $\log_{10}\Sigma_{\Halpha}$ (in unit of $\sfb$) according to the color scale shown above the panels. The top row shows a present-day galaxy extracted from the TNG MW-like galaxy catalogue of \cite{pillepich2023milkywayandromedaanalogs}. The bottom panel shows an ELG from the $z=1$ snapshot of TNG300-1.}
    \label{surface_brightness}
\end{figure*}

\subsection{Sample $\Halpha$ mock galaxies}

Figure~\ref{surface_brightness} displays the $\Halpha$ surface brightness $\Sigma_{\Halpha}$ predicted by our model for two mock galaxies. In each row, the panels display the separate contribution of collisional excitation, recombination, photo-excitation and $\HII$ regions to the $\Halpha$ surface brightness. 

The top row is a $z=0$ galaxy extracted from the TNG50-1 simulation (the galaxy id is 342447), with a total $\Halpha$ luminosity of $L_{\Halpha}=10^{42.55}\ergss$. Although it is classified as a "Milky-Way like" galaxy, it has a relatively high star formation rate, $\text{SFR}=19.07\sfr$. The main contributions to its $\Halpha$ luminosity are CE (46.3\%) followed by HII regions (33.81\%). CE-induced $\Halpha$ emission also traces the corona of hot gas around the galaxy since it can take place even when star formation is absent. The single luminous spot in the $\Sigma_{\Halpha}$ map tracing RR-induced $\Halpha$ emission is an intense burst of star formation, with a SFR larger than $1\sfr$. It is also visible in the contributions from CE, RR and $\HII$ regions.

The bottom row shows $\Sigma_{\Halpha}$ of a galaxy (id 97290) extracted from the $z=1$ snapshot of TNG300-1. Its total $\Halpha$ luminosity is $L_{\Halpha}=10^{44.08}\ergss$ and it has a high star formation rate of  $\text{SFR}=144.9 \sfr$. Despite the presence of an ongoing starburst, CE and RR together contribute 62.2\% of the luminosity. The galaxy also hosts an AGN, which contributes 2.2\% to the total $\Halpha$ luminosity. This satellite galaxy belongs to a cluster of mass $M\simeq 5.6\times 10^{13}\hmsun$. A tail of stripped gas trails the galaxy and produces $\Halpha$ emission which, however, is subdominant compared to the galactic bulge.

\begin{figure}
    \centering
    \includegraphics[width=0.4\textwidth]{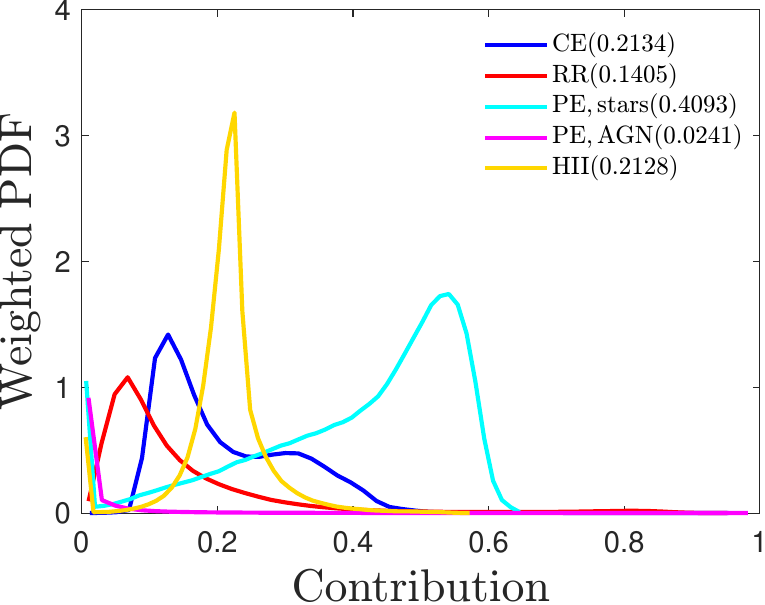}
    \includegraphics[width=0.4\textwidth]{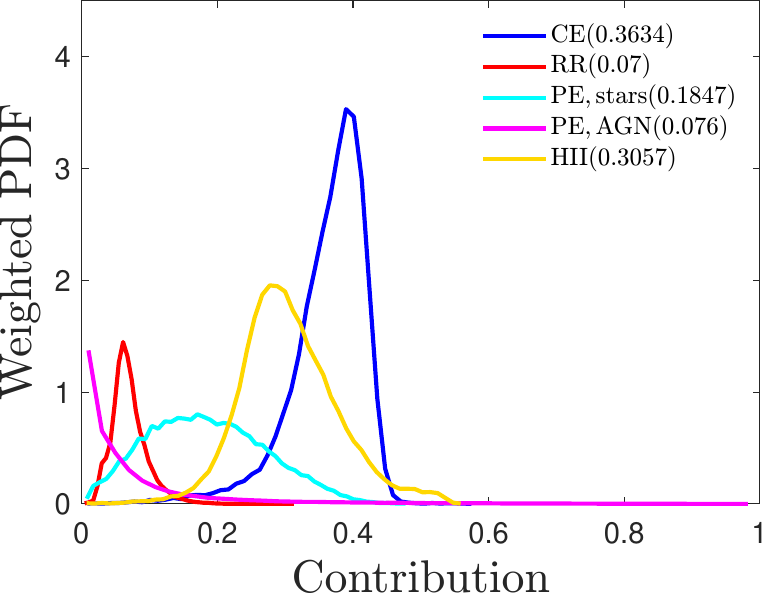}
    \caption{Distributions of the fractional contribution of CE, RR, PE and $\HII$ to the total $\Halpha$ luminosity of an ELG. Results are shown at redshift $z=1$ for all the mock ELGs of the TNG300-1 simulation (top panel), and for those with a luminosity $L_{\Halpha}>10^{42}\ergss$ solely (bottom panel). The area under each distribution is equal to the average, fractional contribution quoted in the insert for each physical process.}
    \label{fig:contribution_PDFs}
\end{figure}

\section{Results}
\label{sec:results}

In this Section, we present predictions of our model for the $\Halpha$ luminosity functions and the HODs of central and satellite galaxies. Additionally, we compare the present predictions to HODs from the \textit{Euclid} Flagship simulations \citep{EuclidFlagship2}.
Unlike \citet{Hadzhiyska_2021} for instance, who developed augmented HOD prescriptions of color-selected mock galaxies that take into account the environment, we will stick to the simplest halo mass-based HOD modelling advocated in \cite{benson/etal:2000,peacock/etal:2001,scoccimarro/etal:2001,berlind/weinberg:2002,zheng/etal:2005}. Extensions of this approach will be considered elsewhere.

\subsection{Properties of the Flagship mocks}
\label{sec:flagship}

The Flagship galaxy mock is a simulated galaxy catalogue designed to support the \textit{Euclid} mission.
The procedure to build it is described in detail in \citet{EuclidFlagship2}; we recap here the relevant steps for our work, which focuses on emission line galaxies with a $\Halpha$ flux $f_{\Halpha}>2\times 10^{-16} \ergsscmm$.

Dark matter halos are populated with a HOD prescription using the \texttt{Scipic} package \citep{Scipic}.
Each halo possesses a central galaxy, while the number of satellites is drawn from a Poisson distribution with mean
\begin{equation}
\big\langle\mathcal{N}_s\big\lvert M \big\rangle = \left(\frac{M}{f M_\mathrm{min}}\right)^\alpha,
\end{equation}
where $M_\mathrm{min}=10^{10} \msun/h$ is the minimum halo mass in the catalog, $f=15$ and $\alpha=1$.

To obtain the $\Halpha$ flux for each galaxy, \texttt{Scipic} goes through the following steps.
First, luminosities in the SDSS $^{0.1}r$ band are assigned to central galaxies to match the luminosity function given by \citet{Dahlen2005}, based on multi-band data taken in the GOODS Survey \citep{Giavalisco2004} and extrapolated up to $z=3$.
A scatter in luminosity is included to adjust the clustering amplitude that would otherwise result too low with respect to observations \citep[e.g.][]{Carretero2015}.
Satellite luminosities are then assigned assuming that in each halo the luminosities follow a Schechter-like probability function.

Galaxies are then assigned a color $^{0.1}(g-r)$, adjusted in order to reproduce the low redshift projected clustering of the SDSS sample \citep{Zehavi2011}.
From this, each galaxy is then assigned a SED, given by a linear combination of templates depending on redshift and color.
Such SEDs do not include emission lines, whose fluxes are instead computed through other physical properties.
First, the galaxy SED is integrated from 1500 to 2300 \r{A} to obtain the unextincted UV flux.
Second, the SFR and the $\Halpha$ flux are computed assuming the relation of \citet{Kennicutt1998} for a Chabrier IMF \citep{Chabrier2003}.
Finally, a Gaussian random scatter of $\Delta\log f_{\Halpha} = 0.05$ is also introduced, resulting in a catalogue of $\Halpha$ emitters with unextincted $\Halpha$ fluxes that can be compared to our mock sample constructed from the TNG simulations.

\subsection{$\Halpha$ Luminosity function}


The $z=1$, 1.5 and 2 snapshots of TNG300-1 contain 1170470, 1516661 and 1813286 ELGs, with $\Halpha$ luminosities $L_{\Halpha}\geq 10^{39}\ergss$ by definition. Fig.~\ref{fig:contribution_PDFs} displays the distributions of the relative contribution of CE, RR, PE and $\HII$ to the total $\Halpha$ luminosity of an ELG. Results are shown at $z=1$ for all the ELGs (top panel) and for those with $L_{\Halpha}\geq 10^{42}\ergss$ (bottom panel). This luminosity cut corresponds to the flux sensitivity of the \textit{Euclid} NISP spectrometer at redshift $z=1$, and reduces the size of the mock ELG sample to 45905 at redshift $z=1$. 
Furthermore, the area under each curve is quoted in the insert. It is equal to the average, fractional contribution of each $\Halpha$ emission channel discussed in Section \S\ref{sec:Haline}. The sum of the areas is thus equal to unity by construction.

For the full sample of mock ELGs, PE from stars and $\HII$ regions dominate the $\Halpha$ luminosity presumably because typical $\Halpha$ emitters do not have hot gaseous atmospheres extending far beyond the stellar distribution.
For ELGs above the \textit{Euclid} flux limit, CE and $\HII$ regions  dominate the $\Halpha$ luminosity, with average fractional contributions of $\sim 36$\% and $\sim 30$\%, respectively. By contrast, the contribution from stellar-induced PE is only $18\%$ of the total $\Halpha$ emission budget. It becomes negligible for the highly luminous galaxies with $L_{\Halpha}\geq 10^{44}\ergss$, whereas the fractional contribution of AGN-induced PE increases with the luminosity threshold to reach 23\% of the $\Halpha$ luminosity when $L_{\Halpha}\geq 10^{44}\ergss$.

\begin{figure}
    \centering
    \includegraphics[width=0.49\textwidth]{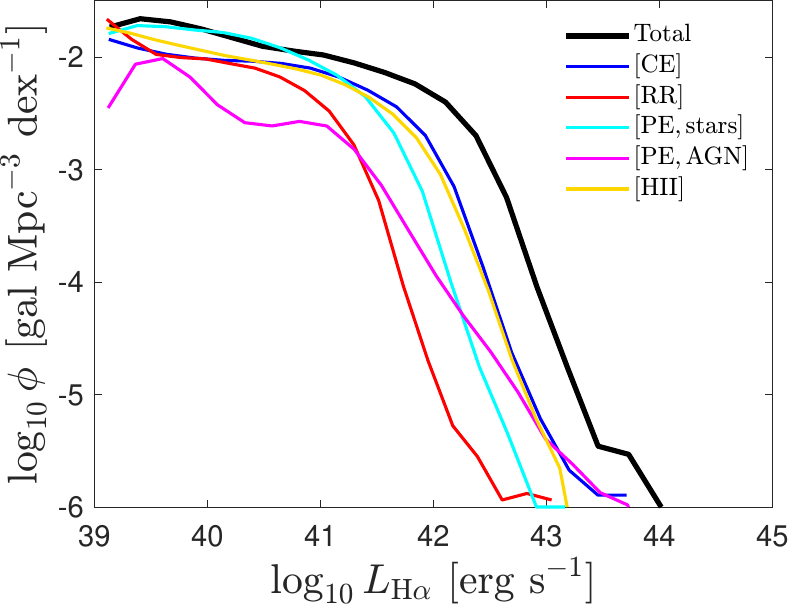}     
    \caption{Luminosity function $\phi(L_{\Halpha})$ of the $z=1$ mock ELGs when each source of $\Halpha$ emission is turned on separately, as indicated in the figure.}
    \label{fig:LF_type_comparison}
\end{figure}

In Fig.~\ref{fig:LF_type_comparison}, the $z=1$ luminosity function (LF) $\phi(L_{\Halpha})$ is shown when each physical mechanism for $\Halpha$ emission is turned on separately. The black curve shows the distribution for the total $\Halpha$ line luminosity, which is better fitted by a broken power-law than a Schechter function. The contributions from AGN and CE dominate at the bright end of $\phi(L_{\Halpha})$. 

\begin{figure*}
    \centering
    \includegraphics[width=0.99\textwidth]{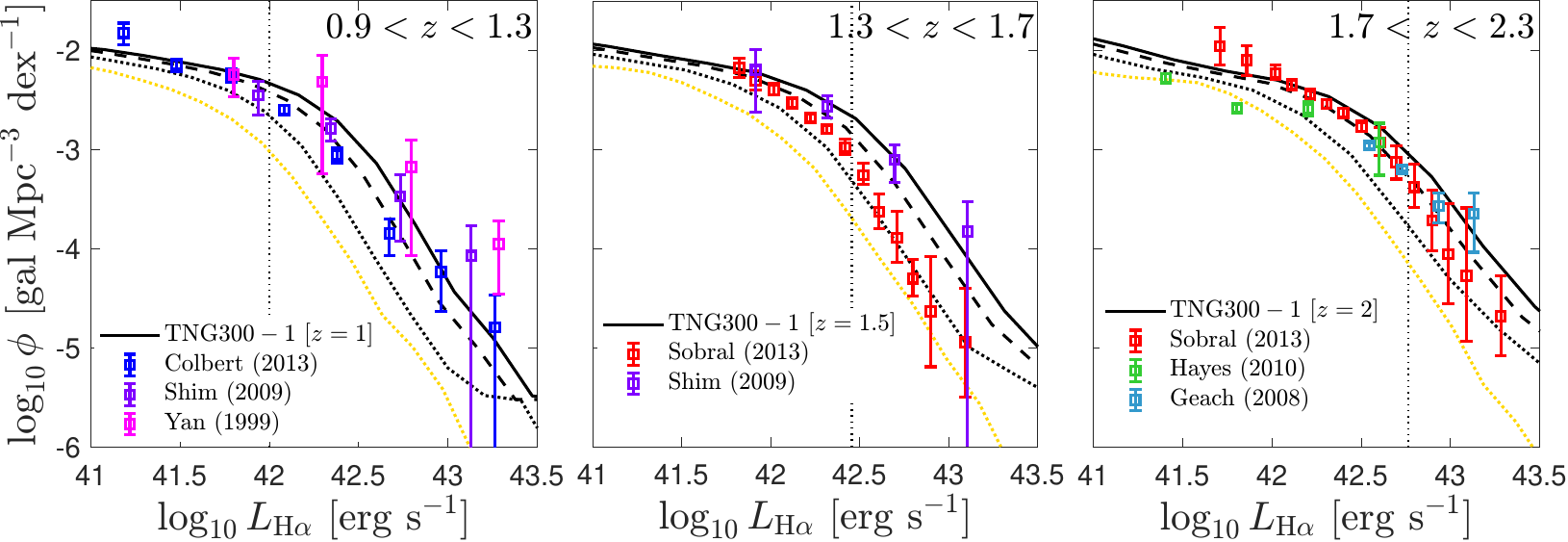}
    \caption{A comparison between the $\Halpha$ luminosity function $\phi(L_{\Halpha})$ obtained from the mock ELGs (curves) and empirical estimates (data points). The theoretical predictions are shown at redshift $z=1$ (left panel), 1.5 (middle panel) and 2 (right panel) as the black curves. The solid curve represents the intrinsic $\Halpha$ LF predicted by our model, 
    while the dashed and dotted lines show the LFs obtained upon applying a uniform extinction of $A_{\Halpha} = 0.3$ and $0.85\ \text{mag}$, respectively. The yellow dotted line shows the contribution of the HII regions with $A_{\Halpha} = 0.85\ \text{mag}$. The empirical estimates of the $\Halpha$ LF are based on observed $\Halpha$ line fluxes in the redshift range quoted in each panel. The thin vertical lines marks the luminosity threshold corresponding to the \textit{Euclid} $\Halpha$ line flux limit of $2\times 10^{-16}\ergsscmm$.}
    \label{fig:LF_with_data}
\end{figure*}

In Fig.~\ref{fig:LF_with_data},  our predictions for the total $\Halpha$ luminosity function $\phi(L_{\Halpha})$ of ELGs are compared to measurements extracted from data at redshift $0.9<z<1.3$ (left panel), $1.3<z<1.7$ (middle panel) and $1.7<z<2.3$ (right panel). These measurements bracket the redshift of the TNG300-1 snapshots considered here. At the low end of the redshift range, the observed LF is based on data from NICMOS on HST \citep{Yan_1999,Shim_2009} as well as WFC3 on HST \citep{Colbert_2013}. At the high end of the redshift range, it is based on narrow-band surveys \citep{Geach_2008, Hayes_2010, Sobral_2013}. Note, however, that we gathered all the data points from \citet{Pozzetti_2016}, who applied some corrections to the original measurements in order to homogenize them. In each panel, the solid curve represents the intrinsic $\Halpha$ LF predicted by our model, which overestimates somewhat the data. The dashed and dotted curve show $\phi(L_{\Halpha})$ after applying to all ELGs a uniform extinction of magnitude $A_{\Halpha}=0.3$ and $0.85\ \text{mag}$, respectively (i.e. all the intrinsic $\Halpha$ luminosities are multiplied by $10^{-0.4A_{\Halpha}}$). This correction takes into account the scattering of $\Halpha$ photons out of the line-of-sight by interstellar dust grains in the source frame. This effect is not corrected for in the data shown in Fig.~\ref{fig:LF_with_data}. The values of $A_{\Halpha}$ quoted above are consistent with \citet{Charlot_2000}. The luminosity functions of the dust-extinguished $\Halpha$ luminosities are in good agreement with the observed LF. Finally, the yellow dotted curve is $\phi(L_{\Halpha})$ extincted by $0.85\ \text{mag}$ when the $\Halpha$ luminosity is given by the empirical $L_{\Halpha}-{\rm SFR}_\text{gal}$ relation (see~\S\ref{sec:HII_regions}). This yields a significant underestimation of the counts in the luminosity range shown here.

As a sanity check, we have computed the cumulative flux count predicted by our model at redshift $z=1$, 1.5 and 2 for a $\Halpha$ line flux limit $2\times 10^{-16}\ergsscmm$ and found they are consistent with empirical estimates \citep{Geach_2010,Pozzetti_2016} and expectations from semi-analytical galaxy formation models \citep{Merson_2017,madar/baugh/shi:2024}. Assuming a uniform extinction of 0.3 mag (which fits better the data points), we find $dN/dz = 10127$, 3161 and $1277\ \text{deg}^{-2}$ at redshift $z=1$, 1.5 and 2, respectively. 

\subsection{A comparison of halo occupation distributions}

We will now compare the mean halo occupation numbers $\Nc$ and $\Ns$ of central and satellite $\Halpha$ emitters extracted from TNG300-1 (or, simply, TNG) using our $\Halpha$ line model (cf.~\S\ref{sec:Haline}) with those of the \textit{Euclid} Flagship simulation (cf.~\S\ref{sec:flagship}). For this purpose, we apply a $\Halpha$ line flux cut of $2\times 10^{-16}\ergsscmm$ to our ELGs, which is identical to $\Halpha$ flux limit of \textit{Euclid}. These mock ELGs are then classified into central and satellite galaxies according to the subhalo identification adopted in the IllustrisTNG simulations. Note also that the mock ELGs used for the comparison do not include dust attenuation in the source frame. 

\begin{figure*}[htp]
    \centering
    \includegraphics[width=1.0\textwidth]{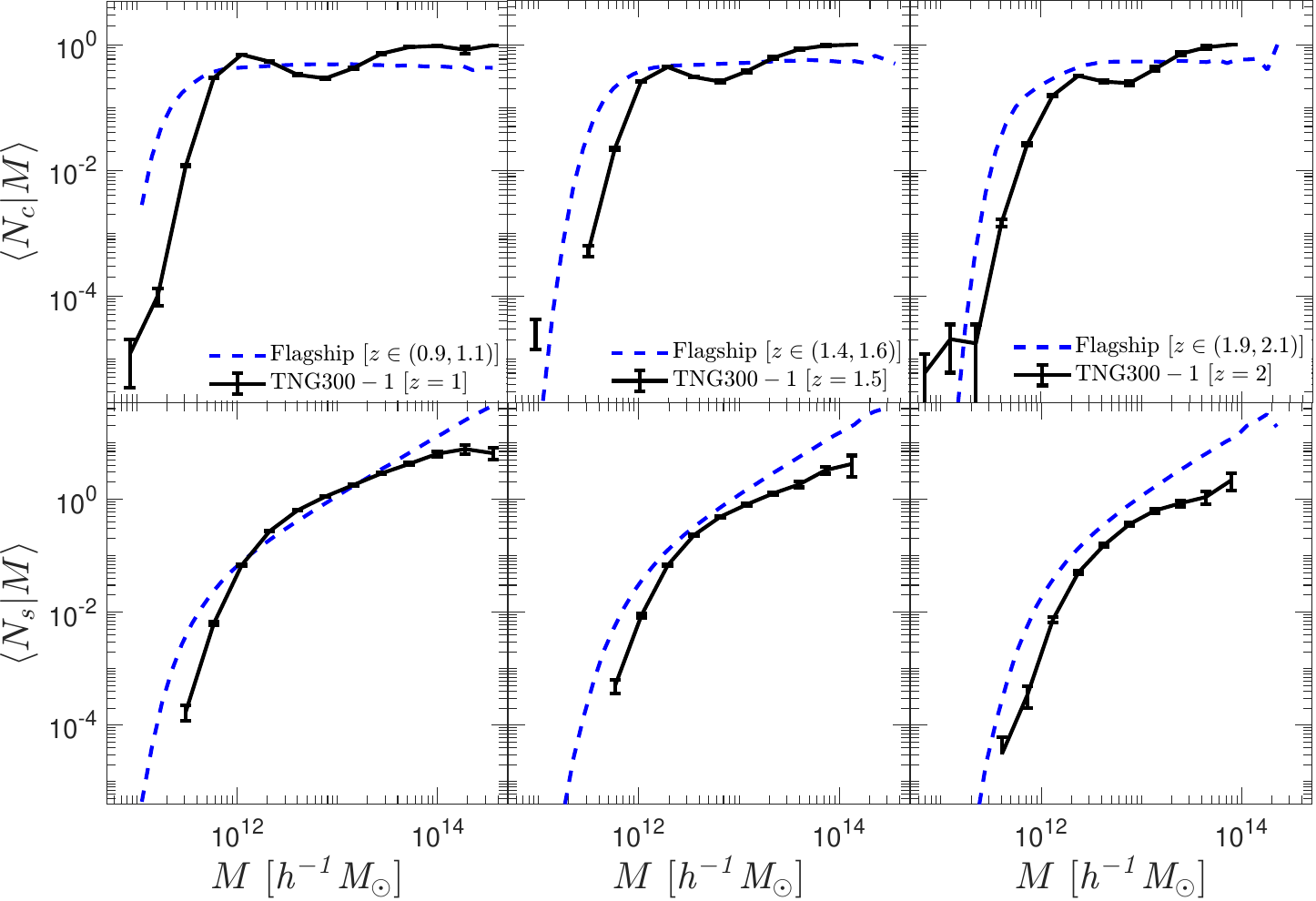}   
    \caption{Mean halo occupation numbers $\Nc$ (top row) and $\Ns$ (bottom row) of central and satellite ELGs with $\Halpha$ line flux above the threshold $2\times 10^{-16}\ergsscmm$. Results are shown at $z=1$ (left panels), 1.5 (middle panels) and 2 (right panel). The predictions of our $\Halpha$ line model applied to TNG300-1 are shown as the (black) solid curves. For comparison, the measurements of $\Nc$ and $\Ns$ from the \textit{Euclid} Flagship mock are shown as the (blue) dashed curves.}
    \label{fig:HOD_main}
\end{figure*} 

The mean halo occupation numbers of the Flagship and TNG ELGs above the Euclid flux limit are shown in Fig.~\ref{fig:HOD_main} at redshift $z=1$ (left panels), 1.5 (middle panels) and 2 (right panels). The top and bottom row show the results for ELGs classified as central and satellite galaxies, respectively. 
The (blue) dashed curves are the central and satellite occupation numbers measured from the Flagship mocks, while the (black) solid curves are $\Nc$ and $\Ns$ for our mock ELG catalogues. The errorbars represent the standard error of the mean, e.g. ${\rm err}\Nc={\rm std}(N_c)/\sqrt{N_h(M)}$ where std is the standard deviation of $N_c$ in a halo mass bin centered on $M$, with $N_h(M)$ being the number of halos in the bin.

While the halo occupation numbers of bright $\Halpha$ emitters predicted by Flagship and TNG are in reasonable agreement, there are two significant differences:
\begin{itemize}
    \item For central galaxies, the TNG HOD exhibit a bump around a halo mass $10^{12}\hmsun$, which is absent in the Flagship HOD. Furthermore, the TNG mean central occupation $\Nc$ converges to unity at the high mass end such that, for $M\gtrsim 10^{13}\hmsun$, almost every halo hosts a central ELG with $L_{\Halpha}\geq 10^{42}\ergss$.
    \item For satellite galaxies, the TNG HOD predicts a shallower rise of $\Ns$ with halo mass, which deviates also from a power law. As a result, the mean satellite occupation $\Ns$ for cluster-like halos with mass $M\geq 10^{14}\hmsun$ is smaller by a factor of few in the TNG mocks. 
\end{itemize}
These deviations persist across the redshift range considered here and, unlike the differences seen in the mean occupations for $M\sim$ a few $10^{11}\hmsun$, do not originate from the different (sub)halo mass definitions adopted by Flagship and TNG. 

The bump in $\Nc$ seen in Fig.~\ref{fig:HOD_main} is also apparent in the HOD of the mock $\Halpha$ emitters presented in \citet{geach/etal:2012,gao/jing/etal:2022}. It is approximately Gaussian, with a characteristic host halo mass $M_\text{eff}\sim 10^{12}\hmsun$ that coincides with the peak of star formation efficiency \citep{behroozi/wechsler/conroy:2013}. Therefore, it is most probably the consequence of the dependence of $L_{\Halpha}$ on SFR and the existence of a characteristic $M_\text{eff}$ for SFR efficiency. Note that such a feature is absent from the central occupation number of optically-selected galaxies, for which a step-like function is a better fit to $N_c$ \citep{zheng/coil/zehavi:2007}. 

\begin{figure*}[htp]
    \centering
    \includegraphics[width=1.0\textwidth]{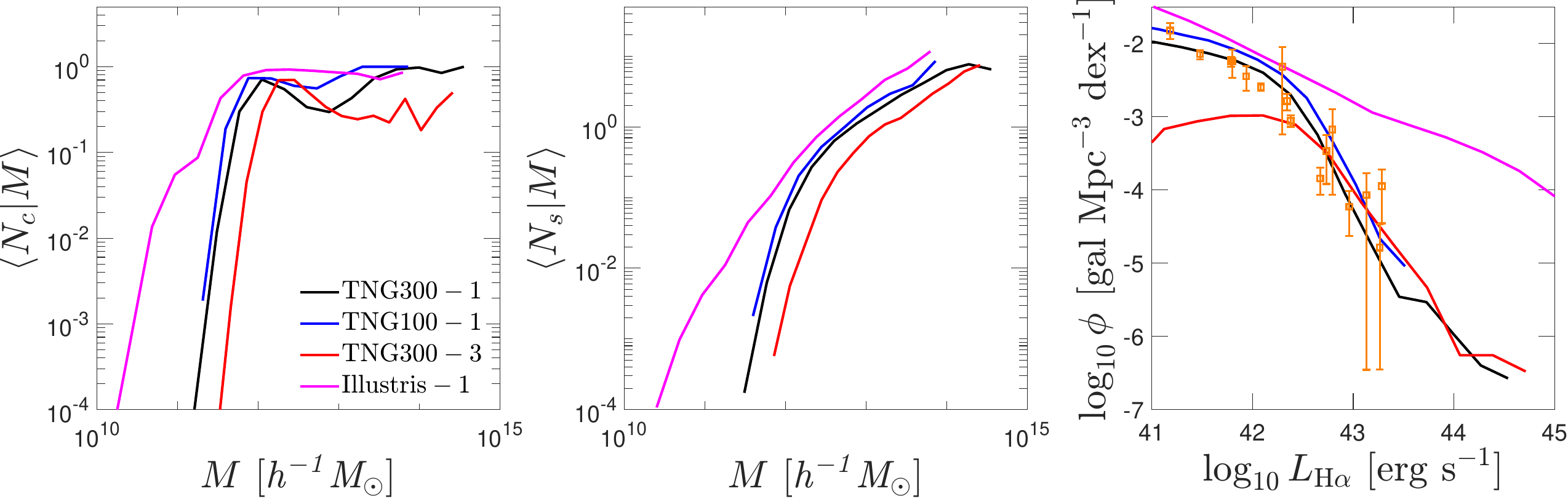}
    \caption{Mean halo occupation numbers $\Nc$ (left panel), $\Ns$ (central panel) and the $\Halpha$ LF (right panel) extracted from the $z=1$ snapshots of simulations listed in Table~\ref{tab:cosmological_settings}. TNG100-1 is the main high-resolution run which is used to calibrate the subgrid model parameters of TNG. TNG300-3 has the same volume as, yet a lower resolution than our fiducial TNG300-1 run. Illustris-1 is an older high resolution simulation of the same cosmological volume as TNG100-1, but with several different prescriptions for feedback compared to the TNG suite of simulations. \chg{The LFs are compared to observed data, plotted in orange (see Fig.~\ref{fig:LF_with_data}).}} 
    \label{fig:simulation_comparison}
\end{figure*}

In contrast to the results of \citet{geach/etal:2012,gao/jing/etal:2022} however, our central occupation number in the redshift range $1\leq z\leq 2$ always asymptotes to unity at high mass, unless we restrict ourselves to luminous ELGs with $L_{\Halpha}\geq 10^{43}\ergss$. This conclusion also holds if the galaxy formation model is changed, or if the $\Halpha$ luminosity is computed solely from the HII regions (see \S\ref{sec:systematics}).

Finally, the relative suppression of $\Ns$ in the TNG mocks at the high mass end may be caused by hydrodynamic effects (such as ram pressure stripping and shock heating) and baryonic feedback (from AGN) in the core of galaxy clusters, which are not modelled in the Flagship mocks (based on pure dark matter N-body simulations). 

\subsection{Model systematics}
\label{sec:systematics}

In this Section, we assess the extent to which variations in the $\Halpha$ emission line model affect the $\Halpha$ luminosity function, and the fraction of central and satellite galaxies above the \textit{Euclid} flux limit. In addition to the galaxy formation model, the main theoretical uncertainties are encapsulated in the parameters $(\lambda_h,r_0,C_{\HII})$.

In Fig.~\ref{fig:simulation_comparison}, we show the effect of changing the simulation volume, numerical resolution and the galaxy formation model on the $z=1$ $\Halpha$ luminosity function and HODs of the galaxies above the \textit{Euclid} flux limit. 
The LFs of the three TNG simulations (whose properties are summarized in Table~\ref{tab:cosmological_settings}) are consistent at high luminosities, where galaxies are large enough for resolution effects to not be important. 
Notwithstanding, the TNG300-1 and TNG300-3 simulations yield satellite occupation numbers of ELGs with $L_{\Halpha}> 10^{42}\ergss$ that differ by at least a factor of a few. Moreover, $\Nc$ asymptotes to a value $<1$ in the high mass limit. TNG100-1 shows consistent HODs with TNG300-1 but with a less pronounced peak of $\Nc$. In addition, it has a slight tendency to overestimate luminosities relative to TNG300-1, which could be a result of resolution effects or cosmic variance. The Illustris-1 simulation has a resolution noticeably better than TNG300-1. However, it implements a different galaxy formation model, such that the $\Halpha$ LF is dominated by $\Halpha$ emission from PE and HII regions at low luminosities, and from AGN-driven PE at high luminosities. As a result, our  Illustris-1 $\Halpha$ emitters mock considerably overestimates the bright end of the $\Halpha$ LF and, thereby, the occupation numbers of central and satellite ELGs with $L_{\Halpha}\gtrsim 10^{42}\ergss$.

\begin{figure*}[htp]
    \centering
    \includegraphics[width=1.0\textwidth]{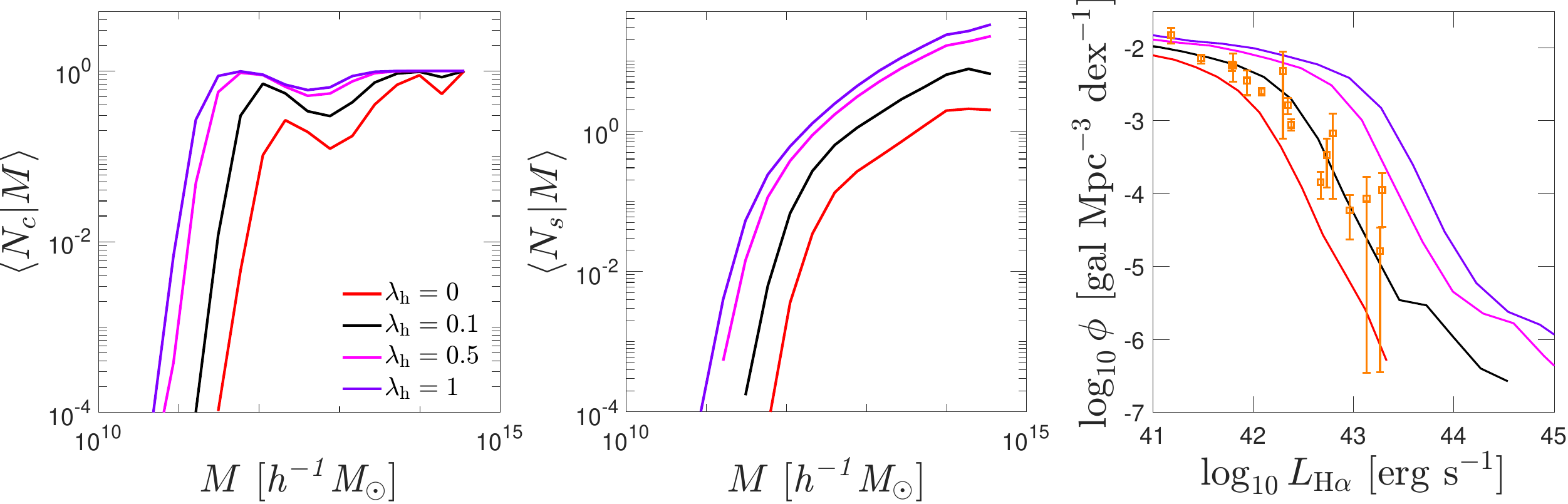}
    \caption{Mean halo occupation numbers $\Nc$ (left panel), $\Ns$ (central panel) and $\Halpha$ luminosity function $\phi(L_{\Halpha}$) (right panel) of the $z=1$ mock ELGs when the fraction $\lambda_\text{h}$ of hot gas in the ISM in star forming cells is varied as indicated in the figure.} 
    \label{fig:NcNs_vary_ISM_frac}
\end{figure*}

\begin{figure*}[htp]
    \centering
    \includegraphics[width=1.0\textwidth]{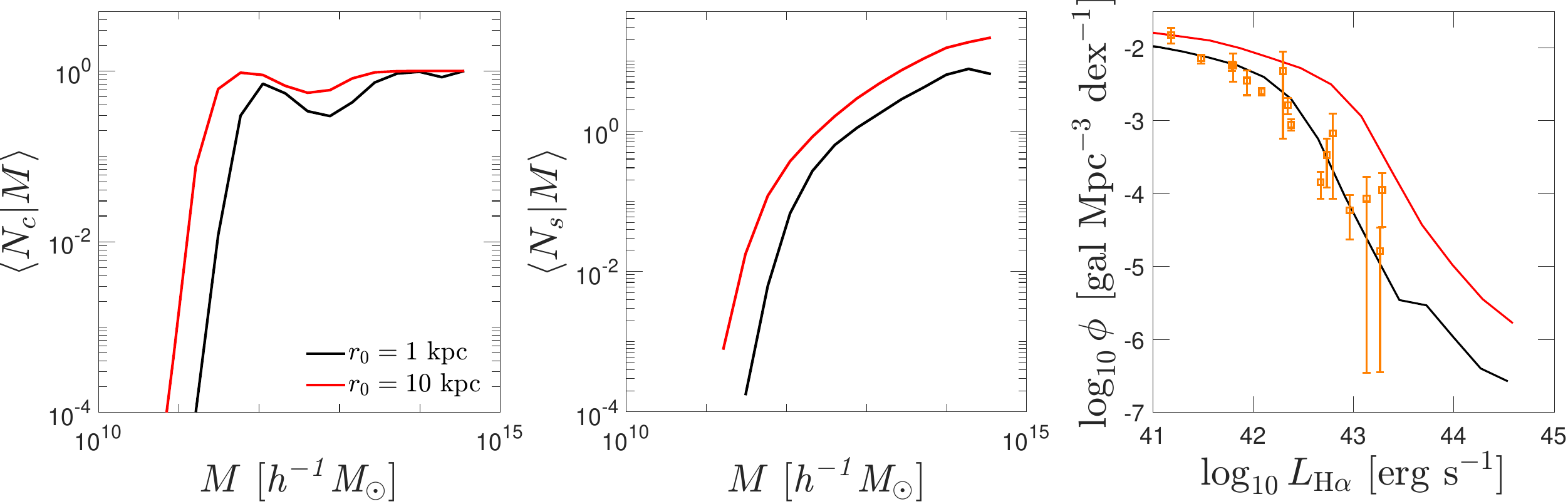}
    \caption{Same as fig.~\ref{fig:NcNs_vary_ISM_frac} for the attenuation length of the stellar radiation in the ISM, which is increased from $r_0=1\kpc$ (fiducial model) to 10$\kpc$.} 
    \label{fig:sys_vary_mfp}
\end{figure*}
 
The fraction $\lambda_\text{h}$ of hot gas in star forming regions is the first key parameter of our $\Halpha$ emission model. It controls the amount of emission arising from the hot, dilute ISM phase in star-forming cells. Although our fiducial choice $\lambda_\text{h}=0.1$ is motivated by the findings of \citet{Springel_2003}, it is prudent to explore how the HOD change when $\lambda_\text{h}$ is varied. 
Fig.~\ref{fig:NcNs_vary_ISM_frac} shows that an increase in the fraction of hot gas from $\lambda_\text{h}=0$ to $\lambda_\text{h}=1$ enhances the bright end of the $\Halpha$ LF considerably. This translates into a fairly uniform increase of $\Ns$ across the whole halo mass range probed by TNG300-1, and a non-vanishing $\Nc$ extending to lower halo mass.  

The attenuation length $r_0$ is the second key parameter, which encapsulates the scattering and absorption of $\Halpha$ photons by dust particles in the ISM of the ELGs. Fig.~\ref{fig:sys_vary_mfp} shows the effect of increasing $r_0$ from its fiducial value $r_0=1\kpc$ to $r_0=10\kpc$, which is a significant fraction of the size of a typical ELG and thus provides likely an upper limit to the actual $r_0$. Increasing $r_0$ has an effect similar to an increase in $\lambda_\text{h}$. The bright end of the $\Halpha$ luminosity function is enhanced, while $\Ns$ as a function of $M$ retains its shape, albeit with a larger normalization.

\begin{figure*}[htp]
    \centering
    \includegraphics[width=1.0\textwidth]{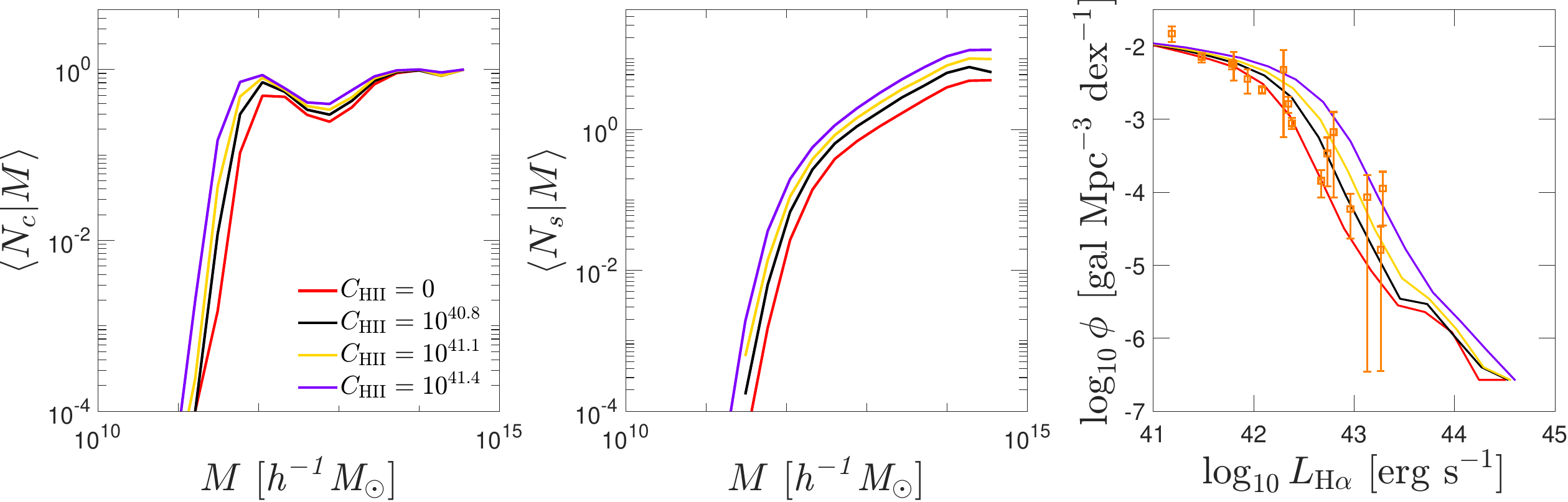}
    \caption{Same as Fig.~\ref{fig:NcNs_vary_ISM_frac} for the proportionality coefficient $C_{\HII}$ of the $\Halpha$ luminosity arising from HII regions (see~\S\ref{sec:HII_regions}). The black curves indicate the fiducial predictions ($C_{\HII}=10^{40.8}$), the red curves show the change when the contribution from HII regions is turned off ($C_{\HII}=0$), the yellow curves are obtained when $C_{\HII}$ is set by the empirical $L_{\Halpha}-\rm{SFR_{gal}}$ relation ($C_{\HII}=10^{41.1}$) while the purple curves  ($C_{\HII}=10^{41.4}$) assume a significantly larger HII contribution.} 
    \label{fig:sys_vary_C_HII}
\end{figure*}

Finally, $C_{\HII}$ determines the importance of the HII region channel of $\Halpha$ emission. Fig.~\ref{fig:sys_vary_C_HII} displays the variations of $\Nc$, $\Ns$ and $\phi(L_{\Halpha})$ relative to the fiducial model with $C_{\HII}=10^{40.8}$ when $C_{\HII}$ is increased to $10^{41.1} - 10^{41.4}$ (which leads to $\Halpha$ luminosities consistent with the empirical SFR - $L_{\Halpha}$ relation only if there are no other sources of $\Halpha$ emission) or set to zero (which implies that $\Halpha$ emission from HII regions is turned off). The resulting variations in the occupation numbers and $\Halpha$ LF show that the other $\Halpha$ emission channels are, at least, equally important for the determination of $L_{\Halpha}$.

\begin{figure}
    \centering
    \includegraphics[width=0.45\textwidth]{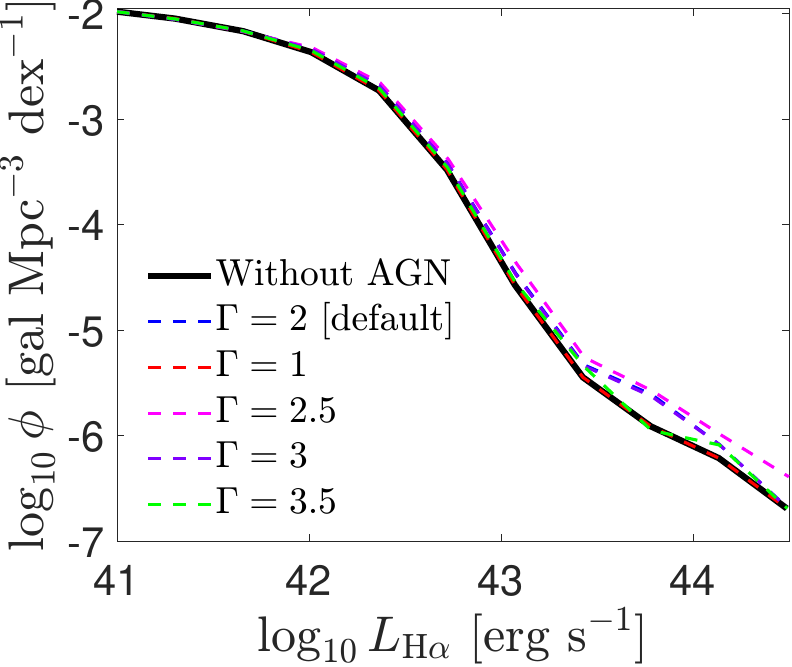}
    \caption{Comparison between the $\Halpha$ luminosity function extracted from the $z=1$ snapshot of TNG300-1 with different implementations of the AGN SED. The black solid curves shows our fiducial LF without the AGN component, while the different dashed curves show the LF with different photon luminosity indices $\Gamma$ as quoted in the figure.}
    \label{fig:LF_AGN_slope_comp}
\end{figure}

We examine the sensitivity of the $\Halpha$ luminosity function to the presence of AGN and their SED power law slope $\Gamma$. As shown in Fig.~\ref{fig:LF_AGN_slope_comp}, the AGN component has an impact on the LF only at high luminosities $L_{\Halpha}>10^{43} \ergss$. Varying $\Gamma$ (at fixed AGN flux $f_{\text{AGN}}$) by a sufficient amount from its fiducial value of $\Gamma=2$ reduces the $\Halpha$ emission induced by AGNs. The impact is largest for $\Gamma=2.5$, for which the total number of ELGs with $L_{\Halpha}>10^{42}\ergss$ increases by 6.8\% (relative to the fiducial model) and 19.4\% (relative to a model without AGNs at all, shown as the solid black curve in Fig.~\ref{fig:LF_AGN_slope_comp}).

To conclude this Section, notice that variations in $(\lambda_\text{h},r_0,C_{\HII})$ affect the slope of the $\Halpha$ LF, which determines the response of the galaxy counts to magnification. For the \textit{Euclid} spectroscopic sample, the systematics in growth rate measurements from magnification bias approaches $1\sigma$ only at the highest redshift \citep{EuclidMagniBiasSpectro}. For our fiducial model parameters, the $z=2$ (logarithmic) slope of the $\Halpha$ LF is $s\simeq -1.73$ at the threshold luminosity, though this value is quite sensitive to the choice of $\lambda_\text{h}$ and $r_0$: we find $s\simeq -2.55$ ($-0.73$) for $\lambda_\text{h}=0$ (0.5) and $s\simeq -0.81$ for $r_0=10\kpc$.
Likewise, the cumulative counts above the \textit{Euclid} flux limit vary noticeably with $(\lambda_\text{h},r_0,C_{\HII})$. For $z=1$ and 0.3 mag extinction for instance, we find $1516\leq dN/dz\leq 18983\ \text{deg}^{-2}$ if $\lambda_\text{h}$ is varied in the range $0\leq \lambda_\text{h}\leq 0.2$, $10127\leq dN/dz\leq 38830\ \text{deg}^{-2}$ for $1\leq r_0\leq 10\kpc$ and $5435\leq dN/dz\leq 14291\ \text{deg}^{-2}$ for $0\leq C_{\HII}\leq 10^{41.1}$. Therefore, it will be interesting to assess whether large variations in the clustering of $\Halpha$ emitters subsist once $(\lambda_\text{h},r_0,C_{\HII})$ are constrained by observed LFs.


\section{conclusions}
\label{sec:conclusions}

We have presented a physically-motivated modelling of spatially-resolved, galactic $\Halpha$ emission that can be applied to hydrodynamical simulations of galaxy formation. In particular, we have attempted to address the lack of a comprehensive theoretical framework for emission from the warm-hot, dilute ISM. We have assumed that the population of the atomic levels are in steady-state and can be computed within the coronal approximation (where excitation occurs either from the ground state or by recombination). These assumptions are justified by the relatively low densities and weak radiation fields in the ISM. We have derived simplified expressions for the $\Halpha$ luminosities induced by collisional excitation, recombination and photo-excitation by stars and AGN, which can be directly related to the physical properties of simulated gas cells or particles. In addition, we have included $\Halpha$ emission from HII regions, which are not spatially resolved in the IllustrisTNG suite of simulations considered here.

We have used our approach to extract mock $\Halpha$ emitters from the IllustrisTNG simulations, which mimic the ELGs surveyed by \textit{Euclid} in the redshift range $1\leq z\leq 2$. We have explored the systematics of our model, and also compared the resulting HOD to that of the Flagship mocks developed by the \textit{Euclid} collaboration \citep{EuclidFlagship2}. Our results can be summarized as follows:
\begin{itemize}
    \item For the bright $\Halpha$ emitters surveyed by \textit{Euclid}, $\Halpha$ emission mechanisms other than HII regions (i.e. collisional excitation, photo-excitation and radiative recombination) contribute an average $\approx 69\%$ of the total $\Halpha$ luminosity of an ELG. This is consistent with expectations drawn from narrow-band imaging surveys, which suggest that the diffuse gas can account for as much as $50\%$ to $70\%$ of the $\Halpha$ emission \citep[see for example][]{Kewley_2019_rev,Zurita_2000}. Specifically, collisional excitation and radiative recombination contribute on average $\approx 40\%$ of the $\Halpha$ luminosity across the redshift range $1\leq z\leq 2$. 
    \item Photo-excitation by UV radiation is often the dominant source powering $\Halpha$ emission in galaxies hosting an AGN, while nearby AGNs can contribute up to a few percent of the total $\Halpha$ emission of a galaxy even when its own AGN is switched off. We stress that this result sensitively depends on the implementation of absorption, AGN radiation geometry etc.
    \item The $\Halpha$ luminosity function predicted by our model is in reasonable agreement with observations after the inclusion of a uniform extinction of $A_{\Halpha}=0.85\ \text{mag}$ consistent with empirical estimations. By contrast, using the SFR-$L_{\Halpha}$ relation inferred from observations at $z\lesssim 1$ yields a $\Halpha$ luminosity function which underestimates the measurements. This shows that the latter cannot be reproduced with a simple SFR-$L_{\Halpha}$ conversion, and other components are necessary.
    \item The mean central occupation number $\Nc$ of our synthetic galaxies converges toward unity at the high (halo) mass end and, moreover, exhibits a local maximum around $M\sim 10^{12}\hmsun$. A similar peak, albeit more pronounced, is also seen in the HODs of other high-redshift ELGs \citep{favole/etal:2016,gao/jing/etal:2022,okumura/etal:2023}.  
    \item The central and satellite occupation numbers predicted by our model are consistent with the Flagship HOD when the key model parameters are set to their fiducial values. However, the HOD of our mock $\Halpha$ emitters above the \textit{Euclid} flux limit are rather sensitive to variations in the fraction of hot gas or the attenuation length of the radiation field in the ISM, and to the details of the baryonic feedback implementation.
\end{itemize}
We have not investigated HOD dependencies beyond the host halo mass. Using a semi-analytical model of galaxy formation, \cite{marinucci/etal:2023} found that, unlike color-selected galaxies, $\Halpha$ emission line galaxies appear to exhibit little assembly bias. However, they considered a single redshift ($z=1$), low values of the host halo mass ($10^{10}\lesssim M\lesssim 10^{12}\msun$), and an emission line model which takes into account the $\Halpha$ emission from $\HII$ regions solely. Therefore, it will be prudent to revisit their results. The model presented here can be straightforwardly applied to assess the importance of assembly bias in ELGs following the separate universe approach advocated by \cite{barreira/etal:2020} for optically-selected galaxies. Moreover, it can be used to quantify the impact of deviations from a Poissonian satellite occupation number and assumptions about the radial distribution of satellite galaxies on the nonlinearity and stochasticity of $\Halpha$ emitters clustering \citep[see e.g.][for related discussions]{cacciato/etal:2012,ginzburg/etal:2017,salcedo/etal:2024}, and their Fingers-of-God (FoG) signature in particular.

Finally, our approach can be extended to other atomic lines such as H$\beta$, $\rm{[OII]\ 3727\AA}$, $\rm{[OIII]\ 5007\AA}$ or $\rm{[SIII]\ 9533\AA}$ for instance. While the H$\alpha$ line remains the best observable out to redshift $z\sim 4 - 5$, a survey like SPHEREx will be able to confidently measure the intensity power spectra of the [OII] and [OIII] lines at $z\lesssim 4$ \citep{Gong_2017}. Furthermore, in the case of the \textit{Euclid} survey, the [OIII] and [SIII] emission lines from interloper galaxies could be misidentified as $\Halpha$ lines \citep{EuclidInterpolers1,EuclidInterpolers2,EuclidInterpolers3}, which would lead to a dilution and a distortion of the clustering signal from the target ($\Halpha$) galaxies \citep[see, e.g.,][]{pullen/etal:2016}. Our model can also be applied to study the effect of line interlopers on the clustering of $\Halpha$ emitters. 
To model those emission lines however, a ionization state model for heavier elements must be developed (the IllustrisTNG simulations provide this information only for hydrogen). Another challenge is the validity of the coronal approximation, which may not be applicable for atoms that have long-lived (metastable) states. 

\section{Acknowledgements}

We thank Shmuel Bialy, Yuval Birnboim, Ari Laor and Adi Nusser for useful discussions.
I.R., V.D. and E.B. acknowledge support from the Israel Science Foundation (grants no. 2562/20 and 1937/19). GP and MC acknowledge support from the Spanish Ministerio de Ciencia, Innovación y Universidades (project PID2021-128989NB).
\texttt{CHIANTI} is a collaborative project involving George Mason University, the University of Michigan (USA), University of Cambridge (UK) and NASA Goddard Space Flight Center (USA). 
This work has made use of CosmoHub, developed by PIC (maintained by IFAE and CIEMAT) in collaboration with ICE-CSIC. It received funding from the Spanish government (MCIN/AEI/10.13039/501100011033), the EU NextGeneration/PRTR (PRTR-C17.I1), and the Generalitat de Catalunya.

\bibliography{references}

\end{document}